\documentclass[preprint,12pt]{elsarticle}

\usepackage{caption}
\usepackage{subcaption}

\usepackage{listings}
\usepackage[table,xcdraw]{xcolor}
\usepackage{comment}

\definecolor{codegreen}{rgb}{0,0.6,0}
\definecolor{codegray}{rgb}{0.5,0.5,0.5}
\definecolor{codepurple}{rgb}{0.58,0,0.82}
\definecolor{backcolour}{rgb}{0.95,0.95,0.92}

\lstdefinestyle{mystyle}{
  backgroundcolor=\color{backcolour}, commentstyle=\color{codegreen},
  keywordstyle=\color{magenta},
  numberstyle=\tiny\color{codegray},
  stringstyle=\color{codepurple},
  basicstyle=\ttfamily\footnotesize,
  breakatwhitespace=false,         
  breaklines=true,                 
  captionpos=b,                    
  keepspaces=true,                 
  numbers=left,                    
  numbersep=5pt,                  
  showspaces=false,                
  showstringspaces=false,
  showtabs=false,                  
  tabsize=2
}

\usepackage{booktabs}

\usepackage{graphicx}
\usepackage{subcaption}
\usepackage{float}
\usepackage{multirow}

\usepackage{hyperref}

\usepackage{amssymb}
\usepackage{nicefrac, xfrac}

\newcounter{bla}

\journal{Computer Physics Communications}

\newcommand{\myof}{SPUMA} %% Name of the software

\begin{document}

\begin{frontmatter}



%\title{\myof{}: a full GPU porting of OPENFOAM\textsuperscript{\textregistered}}

\title{\myof{}: a minimally invasive approach to the GPU porting of OPENFOAM\textsuperscript{\textregistered}}


\author[a]{Simone Bnà\corref{author}}
\author[a,b]{Giuseppe Giaquinto}
\author[a]{Ettore Fadiga}
\author[a]{Tommaso Zanelli}
\author[a]{Francesco Bottau}

\cortext[author] {Corresponding author.\\\textit{E-mail address:} \url{simone.bna@cineca.it}}
\address[a]{Cineca Supercomputing Centre, Via Magnanelli 6/3, 40033 Casalecchio di Reno, Italy}
\address[b]{Dipartimento di Ingegneria Industriale, Università degli Studi di Napoli Federico II, 80125 Napoli, Italy}
%\address[b]{Second Address}

\begin{abstract}
High Performance Computing (HPC) on hybrid clusters represents a significant opportunity for Computational Fluid Dynamics (CFD), especially when modern accelerators are utilized effectively. However, despite the widespread adoption of GPUs, programmability remains a challenge, particularly in open-source contexts. 
In this paper, we present \myof{}\footnote{SPUMA is a CFD software released by Cineca based on OpenFOAM technology. This offering is not approved or endorsed by OpenCFD Limited, producer and distributor of the OpenFOAM software via \url{www.openfoam.com}, and owner of the OPENFOAM\textsuperscript{\textregistered} and OpenCFD\textsuperscript{\textregistered} trade marks.}, a full GPU porting of OPENFOAM\textsuperscript{\textregistered}\footnote{OPENFOAM\textsuperscript{\textregistered} is a registered trade mark of OpenCFD Limited, producer and distributor of the OpenFOAM software via \url{www.openfoam.com}.} targeting NVIDIA and AMD GPUs. The implementation strategy is based on a portable programming model and the adoption of a memory pool manager that leverages the unified memory feature of modern GPUs. This approach is discussed alongside several numerical tests conducted on two pre-exascale clusters in Europe, LUMI and Leonardo, which host AMD MI250X and NVIDIA A100 GPUs, respectively. In the performance analysis section, we present results related to  memory usage profiling and kernel wall-time, the impact of the memory pool, and energy consumption obtained by simulating the well-known DrivAer industrial test case. GPU utilization strongly affects strong scalability results, reaching 65\% efficiency on both LUMI and Leonardo when approaching a load of 8 million cells per GPU. Weak scalability results, obtained on 20 GPUs with the OpenFOAM native multigrid solver, range from 75 \% on Leonardo to 85 \% on LUMI. Notably, efficiency is no lower than 90\% when switching to the NVIDIA AmgX linear algebra solver. Our tests also reveal that one A100 GPU on Leonardo is equivalent 200-300 Intel Sapphire Rapids cores, provided the GPUs are sufficiently oversubscribed (more than 10 million of cells per GPU). Finally, energy consumption is reduced by up to 82\% compared to analogous simulations executed on CPUs.
\\

\end{abstract}
\end{frontmatter}

%% sections

%% ========================================================== %%
% INTRODUCTION
%% ========================================================== %%
\section{Introduction}
\label{sec::introduction}

Over the past few decades, computational fluid dynamics (CFD) has been increasingly applied to a wide range of engineering problems, emerging as an important tool in both industry and research. Today, both commercial and open-source CFD software play an important role in industries such as automotive, aerospace, turbomachinery, wind energy, and air pollutant dispersion, among others.

Industrial applications typically involve complex geometries, turbulent flows, and the need to accurately resolve several time and space scales. Consequently, these simulations often require huge amounts of computational resources, making CFD one of the main actors in the High Performance Computing (HPC) landscape, and also providing challenging tests to the scientific community involved in exascale computing developments.

OpenFOAM \cite{weller_foam} is an open-source, Finite Volume (FV) CFD framework supported by a large international community of users, ranging from academic institutions to private companies.
Designed in the 1990s, OpenFOAM is inherently characterized by a software architecture that targets CPU-based hardware. This prevents it from fully exploiting the tremendous computing power of current hybrid supercomputers.

In high-performance computing, massively parallel architectures based on large numbers of general-purpose CPU cores became the norm in the 1990s, replacing previous vector machines. Graphics processing units (GPUs) were originally developed as specialized hardware to accelerate graphics-related tasks. Over the last two decades, they have played an increasingly important role in accelerating scientific computing applications thanks to their efficiency in handling large amounts of data. 
General-purpose GPU computing has made the current surge in machine learning and AI applications possible. The latter, in particular, is currently the main driving force behind further developments in GPU technology. As a result, GPUs are found in most supercomputers of the Top 500 list \cite{top500}, and cloud computing providers are progressively increasing their offerings of GPU-accelerated computing services.

In the last ten years, many researchers have dedicated efforts to the GPU porting of the OpenFOAM code.
However, most published works have adopted a hybrid approach, consisting in the GPU porting of the linear algebra part of the solution process of the Navier-Stokes equations \cite{Bna2020, zampini2020, Martineau2020, Martineau2021, Olenik2024, cufflink, Dyson2018, Monakov2013, ISPM, Alonazi2018ParCFD2D, SpeedIT, Zahra2011, Rathnayake2017, WILLIAMS2016}. Only a few attempts \cite{Lukasz2012, Lukasz2013, Jasinski2015, suyash24} have focused on the full GPU acceleration of the entire Pressure-Implicit with Splitting of Operator (PISO) and Semi-Implicit Method for Pressure-Linked Equations (SIMPLE) solvers \cite{ferziger02:CMFD}, which are the main algorithms implemented in OpenFOAM to decouple pressure and velocity.

Except for the recent publication by Bnà et al. \cite{Bna2020}, no other GPU development effort has been included in the official OpenFOAM repositories, leading to a plethora of deprecated projects based on old OpenFOAM versions that can be considered obsolete today. An exception is the promising work by Tandon et al. \cite{suyash24}, which demonstrates the renewed willingness of major GPU manufacturers to invest resources into the full porting of OpenFOAM.

The hybrid approach can be easily implemented, but hardly justified in terms of cost-to-solution: OpenFOAM solves the Navier-Stokes equations with a segregated approach, which means that each linear system associated with a PDE has to be copied, converted, and finally solved on the GPUs. The assembly of the matrices is not included in the offloading process and can take up to 50 \% of the entire simulation. In this scenario, due to Amdahl's law, the speed-up cannot be higher than two.
On the other hand, Amdahl's law does not impose any limits on the acceleration of the entire solver. However, the complexity of a production code such as OpenFOAM (consisting of millions of lines of code) makes software integration a daunting task.

The few approaches proposed in the literature that target the entire OpenFOAM workflow are not completely satisfying: Miroslaw et al. \cite{Lukasz2012, Lukasz2013} have fully ported OpenFOAM to GPU using CUDA, while RapidCFD \cite{Jasinski2015} relies on the Thrust library, a CUDA/C++ framework based on the Standard Template Library. The only approach that is portable to several architectures is based on OpenMP target offloading \cite{suyash24}. However, it has been designed for Accelerated Processing Units (APUs) such as the AMD MI300A, which leverage a hardware-unified memory space shared between CPUs and GPUs. As shown in \cite{suyash24}, performance on discrete GPUs suffers a severe degradation due to the overhead of page migrations: more than 65\% of the time is spent updating GPU memory and copying data between host and device. Discrete GPUs are still present in the market and in the supercomputers of the TOP500 list (e.g., Leonardo, Mare Nostrum 5, LUMI), and HPC future hardware trends are not easy to predict.

In this work, we present \myof{} (Simulation Processing on Unified Memory Accelerators), a portable version of OpenFOAM that targets modern heterogeneous hardware. Inspired by the Neko programming model \cite{Jansson2024}, \myof{} supports many modern accelerator backends and does not rely on a single proprietary technology, as was the case in previous studies. The porting strategy described in this paper does not require a major refactoring of the framework, since only minor changes have been implemented in the core classes, reducing the maintainability issues, and it can be applied step-by-step to the entire software.

The paper is divided into 5 sections: in section \ref{sec::openfoam} we briefly introduce OpenFOAM describing the critical aspects of the current architecture; in section \ref{sec::programming_model} we present the programming model implemented in \myof{}, together with inner implementation details, design aspects and portability; in section \ref{sec::performance_evaluation} we validate the software and present the performance of \myof{}, achieved in the pre-exascale supercomputers hosted in Europe (Leonardo and LUMI) running the DrivAer automotive benchmark. Finally, we draw the conclusions in section \ref{sec::conclusions_and_future_work}, together with future work directions.
%% ========================================================== %%

%% ========================================================== %%
% OPENFOAM FRAMEWORK
%% ========================================================== %%
\section{The OpenFOAM framework}
\label{sec::openfoam}

OpenFOAM is a C++ numerical framework that facilitates the development and customization of solvers to perform simulations of fluid flow involving several possible physical phenomena.

With few exceptions, all solvers implemented in OpenFOAM are segregated, which means that every coupled problem (i.e., Navier-Stokes equations) is turned into a sequence of decoupled problems (e.g., PISO, SIMPLE, PIMPLE). Each of them is discretized using the co-located finite volume method (FVM) over unstructured grids. The object-oriented programming features of \textit{C++} are exploited to develop a syntactical model of \textit{equation mimicking} and algebra operations among scalar, vector, and tensor quantities.

Space and time discretization are based on low-order (typically second-order) schemes, and a large set of preconditioners and iterative solvers for the solution of linear systems are available natively.

Finally, the OpenFOAM design is based on message passing parallelism (MPP) to implement the domain decomposition algorithm. MPP is integrated into the framework at a low level, providing high-level APIs that can be used without the need for any parallel-specific coding.

\subsection{Design aspects}

OpenFOAM was designed in the 1990s, a period in which general-purpose accelerators did not yet exist and the amount of memory available was generally low. OpenFOAM uses a lazy approach in the solution process of PDEs; if we do not consider the mesh and the solution fields, every quantity is allocated in heap memory on demand and destroyed as soon as it is no longer needed. Caching of is performed explicitly by the user and is possible only for a limited set of object types. For example, in the main time loop, the linear system and linear solver are created and destroyed at every time step for each scalar equation. Caching these objects as a whole is not allowed, although the user can cache the GAMG structure of the multigrid solver if specified. The process of requesting and releasing memory is governed by the operating system on CPUs (driver on GPUs) and not by the developer. This approach is inefficient on GPUs, as we show in section \ref{sec::performance_evaluation}, but OpenFOAM does not provide more sophisticated memory allocation/deallocation strategies.

An additional important remark concerns the dependence of the OpenFOAM finite volume schemes and linear algebra solvers on a specific data structure represented by the classes \textit{fvMatrix} and \textit{lduMatrix}, addressed in \cite{Bna2020} as \textit{LDUorderedCOO}. 
These classes rely on an implicit degrees of freedom (DOF) map, which means that the mesh data structure reflects the sparsity pattern of the matrix. This approach reduces the memory footprint of the algorithm, but reduces flexibility at the same time, allowing only for scalar problems to be solved. Consequently, the adoption of an alternative matrix data structure without an explicit conversion is impossible without radically changing the entire code. Although this is not the primary focus of the present paper, it is well known that several recent developments \cite{Kreutzer2014, Goumas2009} on matrix formats have focused on maximizing the performance on GPU accelerators and modern hardware in general. %% TODO Aggiungere altri paper?

Since all solvers, preconditioners and smoothers are native and rely on a specific data structure, designed for CPUs, we are able to exploit the runtime selection feature offered by the OpenFOAM framework to include new linear solvers (e.g. algebraic multigrid) optimized for specific hardware. More details are provided in the next sub-section.

%% GPU-MPI aware, Pstream
As mentioned earlier, OpenFOAM adopts a distributed approach to parallelism: the computational mesh is decomposed into several subdomains during the setup phase, and each is assigned to a different parallel process during the simulation.
	All communication between processes is managed by OpenFOAM’s PStream library, which wraps API calls to whichever MPI implementation is available to the user, and is encapsulated into the high-level syntax of the framework.
	This abstraction layer hides all aspects related to parallelization and forces inter-process communication in OpenFOAM to be performed synchronously, favoring readability and maintainability over maximum performance, with the exception of sparse matrix–vector multiplication.
	Asynchronous communication overlapping with computational tasks is a promising avenue for improving OpenFOAM's scalability, on both traditional CPU-based architectures and accelerators, but would require substantial modifications to the code, not in line with the scope of this work.

	On hybrid clusters, GPU-aware implementations of MPI are generally available. These can perform communication directly between GPUs, as long as pointers to GPU memory are provided, without requiring any modification to the API calls. This avoids costly data transfers to host memory.
	Our porting efforts rely on managed memory, a hybrid approach that mirrors allocations on both GPU and host memory. Further details on this are provided in section \ref{lab::unifiedmemory}. Unfortunately, direct GPU-to-GPU communication requires \textit{pure} GPU memory pointers, and while managed memory pointers can be passed to a GPU-aware MPI implementation without modifying the API calls, this triggers implicit transfers between GPU and host memory. These implicit transfers negatively affect the scalability of our code across multiple GPUs.
	One possible solution is to allocate buffers residing exclusively on GPU memory for MPI communication; however, we have not implemented this yet. All results presented in this paper are affected by these implicit memory transfers.
%% ========================================================== %%

%% ========================================================== %%
% PROGRAMMING MODEL
%% ========================================================== %%
\section{Programming model}
\label{sec::programming_model}

OpenFOAM uses C++’s object-oriented features to define a high-level syntax for describing operations and manipulations of fields. For example, operator overloading is used to define algebraic operations between fields, while matrices are built to represent the discrete counterpart of differential operators. This syntax, or Domain Specific Language (DSL), allows for implementing custom solvers, boundary conditions, and additional functionalities with relative ease. One of the main reasons for OpenFOAM’s success and longevity is the versatility provided by its DSL. Another reason is the use of C++ itself, a standardized language with very broad support. This, along with the small number of third-party libraries required by OpenFOAM, has historically ensured its compatibility with a wide range of platforms: the most important prerequisite is a working C++ compiler. 

For GPUs, a vendor-agnostic standardized language like C++ does not yet exist. In the last decade, several approaches have emerged for accelerated GPU programming: proprietary languages such as CUDA \cite{cuda}, open programming languages such as OpenCL \cite{opencl} and HIP \cite{hip}, standard libraries such as stdpar for C++, compiler directive-based APIs such as OpenMP \cite{openmp} and OpenACC \cite{openacc}, external libraries such as Kokkos \cite{Kokkos2014}, the Oxford Parallel DSLs: OP2 \cite{OP22012} and OPS \cite{OPS2014}, OCCA \cite{OCCA2014} and Thrust \cite{thrust}. Each one of these approaches has its advantages and drawbacks. They have varying degrees of support for different GPU vendors (i.e., NVIDIA, AMD, and INTEL) and for the most used programming languages in computational science (i.e., C, C++, and Fortran), with none of them supporting all possible combinations of the above. CUDA, for example, ensures high performance on NVIDIA platforms but lacks support from other vendors. A library like Kokkos, which works as an abstraction layer between the software and the hardware, enables support for multiple platforms but lacks native Fortran support. Since abstraction layers of this kind require specific data types, they either force developers to adopt the same programming language (C++ in the case of Kokkos) or to maintain a mixed-language source code, at considerably more effort. Also, tying core parts of a code to a third-party library brings additional risks, as the library maintainers may decide to stop supporting features crucial to the code. 

When it comes to GPU development, no perfect solution exists. A compromise adopted by Nek5000's developers \cite{Jansson2024} for its GPU-porting is a separation between the high-level interface of the code, which can be maintained in a standardized language (Fortran 2008 in their case), and the low-level numerical backend, for which multiple implementations tailored to specific hardware can be implemented using the most efficient tool available for each one. This approach may be considered sustainable if it reduces as much as possible the size of the hardware-specific portions of the code. If one were to adopt a similar approach in OpenFOAM, the first observation to be made is that its high-level syntax can be left mostly untouched, since explicit loops are rare and mostly contained in overloaded operators for the \textit{Field} and the \textit{GeometricField} classes. However, implementing GPU-ready parallel versions of each of these low-level loops, for several different platforms, is a daunting task that would severely increase the maintenance effort. Section \ref{sec::Portable_programming_abstraction_v2} presents a portable programming abstraction that reduces the number of hardware-specific kernels required by a full porting of OpenFOAM to GPU.

\subsection{Portable programming abstraction}
\label{sec::Portable_programming_abstraction_v2}

OpenFOAM is based on a coarse-grained, distributed memory approach to parallelism: the computational domain is divided into a series of subdomains, and each one is assigned to a different instance of the entire solver. The various solver instances, or processes, communicate through a message passing paradigm at specific points in the solution process, but otherwise operate independently. This approach is effective on CPU-based distributed systems, where each process can run sequentially on a different CPU core.
GPUs are inherently parallel devices, but they implement a shared-memory architecture more adapted to a fine-grained approach to parallelism. The message passing framework that OpenFOAM is based on can still be exploited to divide the computational load over multiple GPUs, but two levels of parallelism are now required. To exploit the parallel nature of each GPU itself, algorithms that would run sequentially on a single CPU core should be adapted for parallel execution. This is a complex task since any loop where the result depends on the order of the operations cannot be trivially parallelized, often requiring the adoption of a different algorithm. 

Another problem emerging from the parallelization of algorithms originally developed for sequential execution is the presence of race conditions: different parallel threads may attempt to access the same memory location at the same time, creating undefined behavior. This is particularly relevant in OpenFOAM, where many algorithms loop over cell faces, accessing memory locations corresponding to each adjacent cell. Two threads may attempt to access the same cell at the same time, resulting in a race condition. This can be prevented using \textit{atomic operations} wherever a race condition may occur. Atomic operations are thread-safe implementations of basic operations that check whether a memory location is in use before accessing it, with some additional overhead. The alternative to atomic operations is the adoption of different algorithms and data structures that avoid the occurrence of race conditions in the first place. In OpenFOAM, this might mean the use of nested loops over the faces of each cell instead of a simple loop over all faces, as explored in \cite{Malenza2022}, and the adoption of a different data structure for matrices (e.g., CSR). This approach, while avoiding the overhead of atomic operations, would require an in-depth refactoring, while this work aims to explore the performance gains that can be achieved by porting OpenFOAM to GPUs without significantly altering its overall structure.

To investigate whether OpenFOAM can be ported effectively to GPU without radical refactoring, Cineca started working on a proof-of-concept reimplementation of some of OpenFOAM’s core functionalities named zeptoFOAM \cite{zepto}. In zeptoFOAM, each parallelizable loop was replaced with a custom-made GPU-executable kernel, while keeping the overall data structure of OpenFOAM. This showed that considerable performance gains can indeed be achieved without a radical refactoring, thanks, in no small part, to the advancements made in recent years by hardware vendors in optimizing atomic operations on GPUs. It also showed that writing a different kernel for each loop (and repeating the process for every target hardware) would be a time-consuming, error-prone, and tedious process. Applying the same approach to the entirety of OpenFOAM would be impractical.

The more efficient approach used in \myof{} was developed observing that, while a wide variety of parallelizable algorithms appear throughout OpenFOAM, they can broadly be divided into three categories:
\begin{enumerate}
	\item{} \textit{for-loop}: the repetition of a block of code N times on N different portions of data (e.g., any algebraic operation between two arrays that is performed for each of their elements).
	\item{} \textit{reduction summation}: an algorithm that operates on one or more arrays to produce a single result as the sum of contributions from each of their elements (e.g., the scalar product returns the sum of products between all the components of two arrays).
	\item{} \textit{reduction comparison}: an algorithm that operates on one array to produce a single result by applying a comparison operation to each of its elements (e.g., min/max functions that return the minimum or maximum value of an array).
\end{enumerate}
It is possible to define an abstract class, named \textit{executor}, that contains interfaces for these three categories of parallelizable algorithms, leaving the operation to be performed N times in parallel as a yet-to-be-defined lambda function (see listing \ref{code::executor_class}). The executor itself is a template class that, using the Curiously Recurring Template Pattern (CRTP) \cite{Abrahams2004}, provides a static interface and accepts one of its derived classes as a template argument. The derived classes of the executor contain hardware-optimized kernels that implement the three categories of parallelizable algorithms defined above. In this way, the amount of hardware-specific code is limited to three kernels in these derived classes, while the rest of the code can call the relevant parallel algorithms through the hardware-agnostic interface provided by the parent \textit{executor} class. The user can then select the desired backend at compile-time.

\begin{lstlisting}[language=C++, caption=Application of the CRT pattern to the \textit{executor} class., label=code::executor_class]
	namespace Foam
	{
		template <typename T>
		class executor
		{
			public:
			executor() = default;
			~executor() = default;
			
			// No copy constructor
			executor(executor<T> &other) = delete;
			
			template<typename F> 
			void parallelFor(F& lambda, const label& size)
			{
				(static_cast<T*>(this))->_backendFor(lambda, size);
			};
			
			template<typename F, typename resultT>
			void reductionSum(F& lambda, resultT* result, const label& size)
			{
				(static_cast<T*>(this))->_backendReductionSum(lambda, result, size);
			};
			
			template<typename F,typename Op, typename resultT>
			void reductionCompare(F& lambda,Op& op, resultT* result, const label& size)
			{
				(static_cast<T*>(this))->_backendReductionCompare(lambda,op, result, size);
			};
		}
		
		...
		
		class cudaExecutor
		: 
		public executor<cudaExecutor>
		{
			public:
			template<typename F>
			void _backendFor(F& lambda, const label& size);
			
			template<typename F, typename resultT>
			void _backendReductionSum(F& lambda, resultT* const __restrict__ result, const label& size);
			
			template<typename F,typename Op, typename resultT>
			void _backendReductionCompare(F& lambda, Op& op, resultT* const __restrict__ result, const label& size);
		};
	\end{lstlisting}
	
	Once this executor infrastructure is set up, the porting of OpenFOAM to accelerators can be realized progressively by looking for portions of code that can be parallelized and wrapping them in a lambda function, which is then passed to the parallel executor corresponding to its category (e.g. \textit{parallelFor}). As an example, listing \ref{code::gpu_mag} shows an application of this concept to the \textit{mag} function.
	
	\begin{lstlisting}[language=C++, caption=Porting to GPU of the \textit{mag} function., label=code::gpu_mag]
		template<class Type>
		void mag
		(
		Field<typename typeOfMag<Type>::type>& result,
		const UList<Type>& f1
		)
		{
			typedef typename typeOfMag<Type>::type resultType;
			if (result.usePool() && f1.usePool())
			{
				/* Check fields have same size */
				checkFields(result, f1, "f1 = mag(f2)");
				auto rp = result.begin();
				auto f1p = f1.cbegin();
				auto Lambda = [=](label i){rp[i] = mag(f1p[i]);};
				foamExecutor exec;
				exec.parallelFor(Lambda,result.size());
			}
		}
	\end{lstlisting}
	
	\subsection{Memory pool}
	\label{lab::memorypool}
	
	An efficient memory management on NUMA and GPU architectures is crucial when the amount of memory is limited. Memory pools are algorithms that rely on pre-allocated blocks of memory to efficiently manage dynamic memory allocation. Pools avoid memory fragmentation and increase performance by reusing the allocated memory \cite{umpire2020, suyash24}.
	
	As we reported in section \ref{sec::openfoam}, OpenFOAM uses a lot of temporary arrays that require allocation and deallocation in a short period of time: memory pools are a valid tool to reduce the overhead associated with these operations. The impact of memory pools is further examined in Section \ref{subsec::profiling_results}, where different allocation strategies are compared. 
	
	We defined a semi-automatic policy to determine which arrays are allocated in the pool: every \textit{Field} object, or instance of a class derived from \textit{Field}, is automatically allocated in the pool, while for \textit{List} objects this must be specified manually. This strategy avoids cluttering the memory pool with small objects that do not participate in GPU-accelerated computations.
	Since memory allocation mechanisms differ across hardware vendors, we developed a small interface that exposes a few APIs to several hardware-specific backends. On top of this interface, we implemented three algorithms to orchestrate the allocation and deallocation of several memory blocks: the \textit{dummy pool}, the \textit{fixed-size pool}, and the \textit{dynamic-size pool}.
	
\begin{figure}
		\centering
		\includegraphics[width=0.8\linewidth]{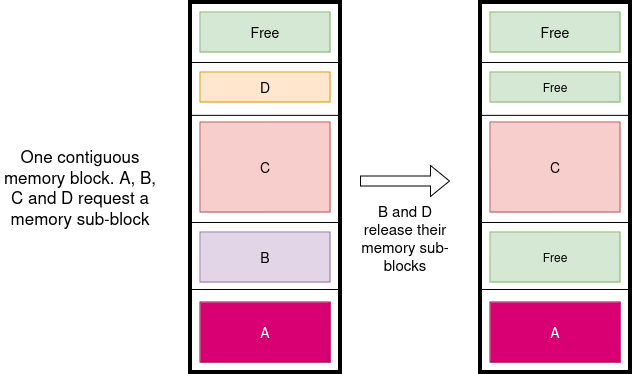}
		\caption{\textit{Fixed-size} pool: one contiguous memory block of fixed size is divided in many sub-blocks; allocation and deallocation of a new memory space involves only pointer handling.}
		\label{fig::memory_pool}
	\end{figure}

	The \textit{dummy pool}, useful for debugging and profiling, allocates resources in a standard way, using the vendor APIs. The \textit{fixed-size pool} is a slightly more sophisticated algorithm, where a single block of constant size and large enough to contain all the fields used during the simulation is allocated through the vendor APIs. 
	Whenever the program asks for a new memory space, the pool manager checks if the required amount of memory is available. If the answer is positive, no allocation is performed, but a pointer to a position in the pool is stored in a list and returned. At the same time, the available size is reduced accordingly. If such an amount of memory is not available, a fatal error stops the simulation. When a memory space is no longer required, the pool manager does not perform any deallocation, it simply deletes the pointer from the list and increases the amount of available space. This mechanism is visually represented in Figure \ref{fig::memory_pool}.
	
	As a third option, more complicated pooling strategies to manage \textit{dynamic-size pools} are contained in the \textit{Umpire} library. \textit{Dynamic-size pools} do not require evaluating the maximum memory needed for the simulation; instead, only an estimate is enough.  In this work, the only original contribution regarding \textit{Umpire} is the development of an interface between \myof{} and this third-party library. We refer to \cite{umpire2020} for more details on the implementation side.
	
	%%%%%%%%%%%%%%%%%%%%%%%%%%%%%%%%%%%%%%%%%%%%%%%%%%%%%%%%%%%%%%%%%%%%%%%%%%%

	%%%%%%%%%%%%%%%%%%%%%%%%%%%%%%%%%%%%%%%%%%%%%%%%%%%%%%%%%%%%%%%
	
	\subsection{Unified memory}
	\label{lab::unifiedmemory}
	
	Unified memory (or managed memory) is a vendor technology that allows applications to allocate data that can be accessed or modified from code running on either CPUs or GPUs. When the application is accessing data allocated on managed memory, the driver and/or the hardware take care of migrating the missing memory pages to the memory of the accessing device.
	
	If unified memory is combined with a pool strategy, allocation and data migration are completely transparent to the developer, considerably reducing the effort spent in the porting process. The approach described in section \ref{sec::Portable_programming_abstraction_v2} owes its efficiency to the fact that the developer can offload to the GPU different portions of the code incrementally, without caring about explicit memory management. Since the data can be moved from CPUs to GPUs automatically, the developer can focus only on the verification of the results once a new algorithm is ported to the accelerator.
	
	The porting to GPUs is completed as soon as data processed by GPUs are not moved back and forth during the simulation. At the end of this process, only a small amount of data is copied from GPU to CPU, e.g., scalar parameters that are printed on screen or used to determine whether to proceed or not in the simulation workflow. Intensive data movement only remains present at the beginning of the simulation due to the offloading of mesh quantities and solution fields, read by CPUs from disk, and other I/O operations.

	%%%%%%%%%%%%%%%%%%%%%%%%%%%%%%%%%%%%%%%%%%%%%%%%%%%%%%%%%%%%%%%
	
	\subsection{External solvers: amgx4foam}
	\label{lab::amgx4section}
	
	The development of GPU-specific linear algebra features (e.g., preconditioners, smoothers, and multigrid algorithms like AMGs) can be seen as a further step in the GPU porting strategy presented in this work. However, such algorithms are necessary to solve CFD problems, and the absence of an efficient implementation complicates the evaluation of the actual benefits of accelerating the numerical solution of the Navier-Stokes equations with GPUs.
	
	One of the solutions presented is linking \myof{} with third-party libraries implementing state-of-the-art linear solvers on GPUs. In particular, NVIDIA maintains a vendor-optimized library targeting distributed algebraic multigrid algorithms (among other linear preconditioners and solvers) \cite{Naumov2015}. This library, called AmgX, can be easily plugged into \myof{} by exploiting the runtime selection features offered by the OpenFOAM framework, as highlighted in section \ref{sec::openfoam}.
	
	A side product of this work is a new implementation of \textit{amgx4foam}, which is an external library that extends the number of linear solvers and preconditioners that are natively ported to GPUs in \myof{}. This library is a refactoring of previous developments, described in \cite{Bna2020, Martineau2020, Martineau2021}. Contrary to the hybrid approaches described in these studies, in this case, the matrices already reside in the GPU memory and no offloading from CPUs to accelerators is performed. Therefore, only a conversion from the OpenFOAM native format to the CSR format is now required. The conversion is implemented using the same strategy proposed in \cite{Martineau2020, Martineau2021}, based on a map computed by the Radix sort algorithm applied to the OpenFOAM arrays representing the sparsity pattern of the matrix. The same methodology can be applied to other external linear algebra libraries such as HYPRE \cite{Hypre2002}, PETSc \cite{petscgpu-mills2021}, Ginkgo \cite{Ginkgo2022}, rocALUTION \cite{rocalution}, Paralution \cite{paralution}, BootCMatchG (\cite{BootCMatchG-BERNASCHI2020}), Chronos \cite{Chronos-Janna2021}, SParSH-AMG \cite{SParSH-AMG-2020}---all implementing a GPU-accelerated algebraic multigrid linear solver such as HYPRE's BoomerAMG and PETSc's GAMG---to extend the list of GPU-ready linear solvers available in \myof{}.
%% ========================================================== %%

%% ========================================================== %%
% PERFORMANCE EVALUATION
%% ========================================================== %%
\section{Performance evaluation}
\label{sec::performance_evaluation}

\subsection{Experimental setup} 
\label{sec::experimental_setup}

The numerical experiments presented in this section have been conducted with \myof{} and OpenFOAM-v2412 on Leonardo and LUMI, two European pre-exascale supercomputers that respectively host NVIDIA A100 and AMD MI250X GPU cards. The hardware characteristics of the two clusters are reported in Tables \ref{tab:leonardo_hardware} and \ref{tab:lumi_hardware}. As reported in Table \ref{tab:lumi_hardware} AMD MI250X GPU cards contain two Graphics Compute Dies (GCDs), each functioning as a discrete GPU. In the remainder of the paper, references to AMD GPUs pertain to the individual GCDs.
The software stacks are reported in Table \ref{tab:software_stack}.

\begin{figure}[h]
	\centering
	%% First subfigure
	\begin{subfigure}[b]{0.48\textwidth}
		\centering
		\includegraphics[width=7cm]{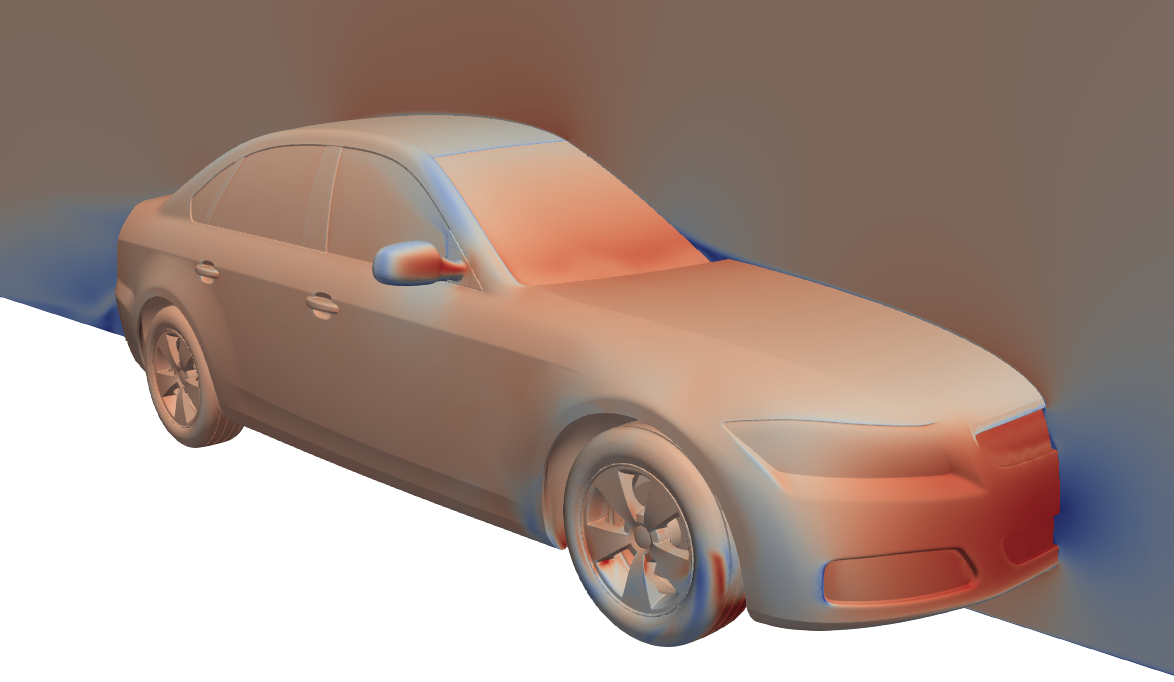}
		\caption{}
		\label{fig:notchback}
	\end{subfigure}
	\hfill
	% Second Subfigure
	\begin{subfigure}[b]{0.48\textwidth}
		\centering
		\includegraphics[width=7cm]{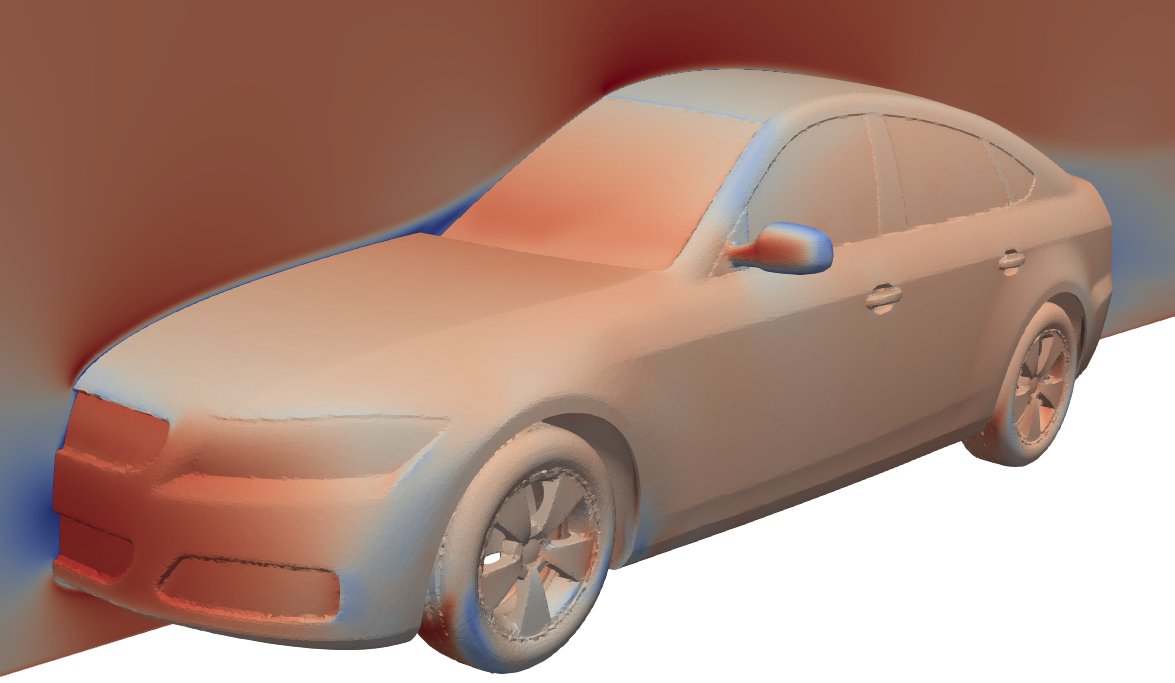}
		\caption{}
		\label{fig:fastback}
	\end{subfigure}
	\caption{DrivAer model in the notchback (a) and fastback (b) configuration.}
	\label{fig:drivaer-configurations}
\end{figure}

Validation, profiling, and performance tests have been performed using the DrivAer industrial test case, a popular automotive benchmark described in \cite{drivAer2012}. The vehicle has been simulated in two configurations, the fastback geometry and the open-closed cooling version (see Figure \ref{fig:drivaer-configurations}). Table \ref{tab:drivaer-meshes} reports the acronyms used in the
paper associated with the different meshes and configurations used in the simulations. The complete setup of the two test cases is included in references \cite{fastback_drivaer} and \cite{occ_drivaer}, associated with the fastback version and the open-closed cooling version respectively. In both cases, the original problem was simplified by adopting static wheels and static ground.

Throughout the paper, CPU-only runs were performed in the CPU cluster partition (\text{i.e.} Leonardo-DCGP or LUMI-C) while GPU runs were executed in the GPU cluster partition (\text{i.e.} Leonardo-Booster or LUMI-G).

%%%%%%%%%%%%%%%%%%%%%%%%%%%%%%%%%%%%%%%%%%%%%%%%%%%%%%%%%%

%%%%%%%%%%%%%%%%%%%%%%%%%%%%%%%%%%%%%%%%%%%%%%%%%%%%%%%%%%

%%%%%%%%%%%%%%%%%%%%%%%%%%%%%%%%%%%%%%%%%%%%%%%%%%%%%%%%%%
\begin{center}
	\begin{table}[h!]
		\resizebox{1.1\textwidth}{!}
		{
			\begin{tabular}{lcc}
				\toprule
				Partition/spec & Booster & DCGP \\
				\midrule
				Model & \begin{tabular}{@{}c@{}}Atos BullSequana X2135 "Da Vinci" \\ single-node GPU blade\end{tabular}  & \begin{tabular}{@{}c@{}}Atos BullSequana X2140 \\ three-node CPU blade\end{tabular}  \\
				%Racks & 116 & 22 \\
				Nodes & 3456 single socket 32 cores & 1536 dual socket 56 cores \\
				Processors & \begin{tabular}{@{}c@{}}Intel Ice Lake CPU\\1 x Intel Xeon Platinum 8358, \\ 2.60GHz TDP 250W\end{tabular} & \begin{tabular}{@{}c@{}}Intel Sapphire Rapids CPU\\2 x Intel Xeon Platinum 8480+, \\ 2.00 GHz TDP 350W\end{tabular} \\
				Accelerators & \begin{tabular}{@{}c@{}}4 x NVIDIA A100 DaVinci GPUs/node\\ 64GB HBM2e (1640 GB/s) NVLink 3.0 (200GB/s)\end{tabular} & - \\
				Cores & 32 cores/node & 112 cores/node \\
				RAM & 512 (8x64) GB DDR4 3200 MHz & 512 (16 x 32) GB DDR5 4800 MHz \\
				Peak performance & 306.31 PFlop/s & 9.43 PFlop/s  \\
				Internal network & 2 x dual port HDR100 per node & single port HDR100 per node \\
				\bottomrule
			\end{tabular}
		}
		\caption{\label{tab:leonardo_hardware} Leonardo hardware characteristics.}
		\bigskip
		\resizebox{1.1\textwidth}{!}{
			\begin{tabular}{lcc}
				\toprule
				Partition/spec & LUMI-G & LUMI-C \\
				\midrule
				Nodes & 2978 single socket 64 cores & 2048 dual socket 64 cores \\
				Processors & \begin{tabular}{@{}c@{}}1 x AMD EPYC 7A53 CPU (Trento), \\ 3.5GHz \end{tabular} & \begin{tabular}{@{}c@{}}2 x AMD EPYC 7763 CPU (Milan), \\ 2.45 GHz base, 3.5 GHz boost \end{tabular} \\
				Accelerators & \begin{tabular}{@{}c@{}}4 x AMD MI250X GPUs\\ 2nd Gen AMD CDNA architecture\\ 2 x 64GB HBM2e (3200 GB/s) per GPU\end{tabular} & - \\
				Cores & 64 cores/node & 128 cores/node \\
				RAM & 512 (8x64) GB DDR4 & 256 (8 x 32) GB (or higher)\\
				Peak performance & 531.51 PFlop/s  & 7.63 PFlop/s  \\ %% TOT: 379.70 Pflop/s
				Internal network & \multicolumn{2}{c}{1 x 200 GB/s HPE Cray Slingshot-11 interconnect} \\
				\bottomrule
			\end{tabular}
		}
		\caption{\label{tab:lumi_hardware} LUMI hardware characteristics.}
	\end{table}
\end{center}
%%%%%%%%%%%%%%%%%%%%%%%%%%%%%%%%%%%%%%%%%%%%%%%%%%%%%%%%%%

%%%%%%%%%%%%%%%%%%%%%%%%%%%%%%%%%%%%%%%%%%%%%%%%%%%%%%%%%%
\begin{table}[h!]
	\centering
	%\resizebox{1.\textwidth}{!}
	%{
		\begin{tabular}{c c c}
			\hline
			Cluster&Package & Version \\
			\hline
			&CUDA & 12.1 \\
			Leonardo&NVCC Compiler & 23.11 \\
			(Intel + NVIDIA)&OpenMPI & 4.1.6 \\
			&AmgX & 2.5 \\
			\hline
			&HIP Compiler (ROCm)& 6.0.3 \\
			%ROCm & 6.0.3 \\
			LUMI (AMD)&AMD clang Compiler& 17.0.0 \\
			&Cray-MPICH & 8.1.29 \\
			\hline
		\end{tabular}
		%}
	\caption{Software stack.}
	\label{tab:software_stack}
\end{table}
%%%%%%%%%%%%%%%%%%%%%%%%%%%%%%%%%%%%%%%%%%%%%%%%%%%%%%%%%%

%%%%%%%%%%%%%%%%%%%%%%%%%%%%%%%%%%%%%%%%%%%%%%%%%%%%%%%%%%
\begin{table}[ht]
	\centering
	%\resizebox{1.\textwidth}{!}
	%{
		\begin{tabular}{c c c c}
			\hline
			Name  & Configuration & \# of Cells & \# of Faces\\
			\hline
			DrivAer22M      & fastback            & 22.555.718   & 68.559.955 \\
			occDrivAer65M   & open-closed cooling notching & 65.334.765   & 203.271.868 \\
			occDrivAer120M  & open-closed cooling notching & 120.899.906  & 374.098.306  \\
			occDrivAer235M  & open-closed cooling notching & 235.741.179  & 726.263.835 \\
			\hline
		\end{tabular}
		%}
	\caption{\label{tab:drivaer-meshes}DrivAer configurations: DrivAer22M is taken from \cite{fastback_drivaer} while occDrivAer configurations are taken from \cite{occ_drivaer}.}
	\label{tab:meshes}
\end{table}
%%%%%%%%%%%%%%%%%%%%%%%%%%%%%%%%%%%%%%%%%%%%%%%%%%%%%%%%%%

\subsection{Validation results}
\label{subsec::verification_results}
An important requirement for \myof{} is the ability to replicate the results of its CPU reference, OpenFOAM-v2412.
The RANS DrivAer22M test case was adopted to validate the results, using the setup reported in \ref{appendix:simulation_setup}. Simulations were run with \myof{} and OpenFOAM-v2412 for 100 SIMPLE iterations, starting from the flow field at Time=2000 obtained with OpenFOAM-v2412.

%%%%%%%%%%%%%%%%%%%%%%%%%%%%%%%%%%%%%%%%%%%%%%%%%%%%%%%%%%

\begin{table}[ht]
	\centering
	\begin{tabular}{c c c c c c}
		\hline
		& \multicolumn{2}{c}{$C_d$} & \multicolumn{2}{c}{$C_l$}   \\
		& Leonardo & LUMI & Leonardo & LUMI \\
		\hline
		OpenFOAM-v2412 (GAMG) & 0.28498 & 0.28498 & 0.01816 &0.01817 \\
		\myof{} (GAMG)        & 0.28498 & 0.28498 & 0.01816 & 0.01816 \\
		\myof{} (AMGX)        & 0.28498 & - & 0.01816 & - \\
		\hline
		OpenFOAM-v2412 (PCG)  & 0.28498 & 0.28498 & 0.01817 & 0.01817 \\
		\myof{} (PCG)         & 0.28498 & 0.28498 & 0.01817 & 0.01817 \\
		\hline
	\end{tabular}
	\caption{DrivAer22M. Mean drag and lift coefficients using different solvers for the pressure equation. The average procedure is performed in the interval [$T_i=2000$, $T_f=2100$].}
	\label{tab:mean_force_coeff_T_2000}
\end{table}

%%%%%%%%%%%%%%%%%%%%%%%%%%%%%%%%%%%%%%%%%%%%%%%%%%%%%%%%%%

\begin{figure}[ht]
	\centering
	%% U
	\begin{subfigure}[b]{0.48\textwidth}
		\centering
		\includegraphics[width=\textwidth]{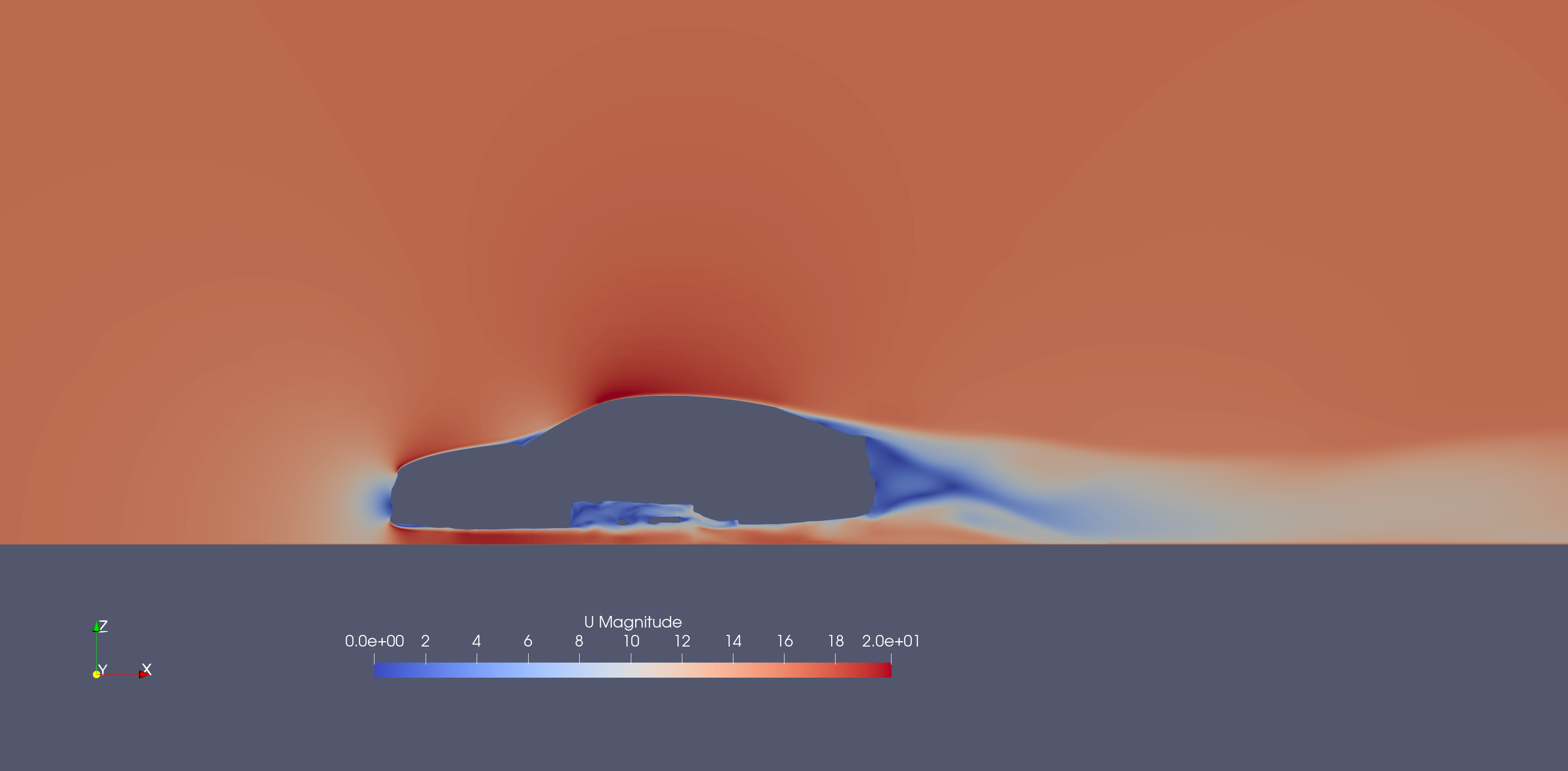} %% img/validation/snapshots/
		\caption{U-\myof{}}
		\label{fig:U-sub1}
	\end{subfigure}
	\hfill
	% Second Subfigure
	\begin{subfigure}[b]{0.48\textwidth}
		\centering
		\includegraphics[width=\textwidth]{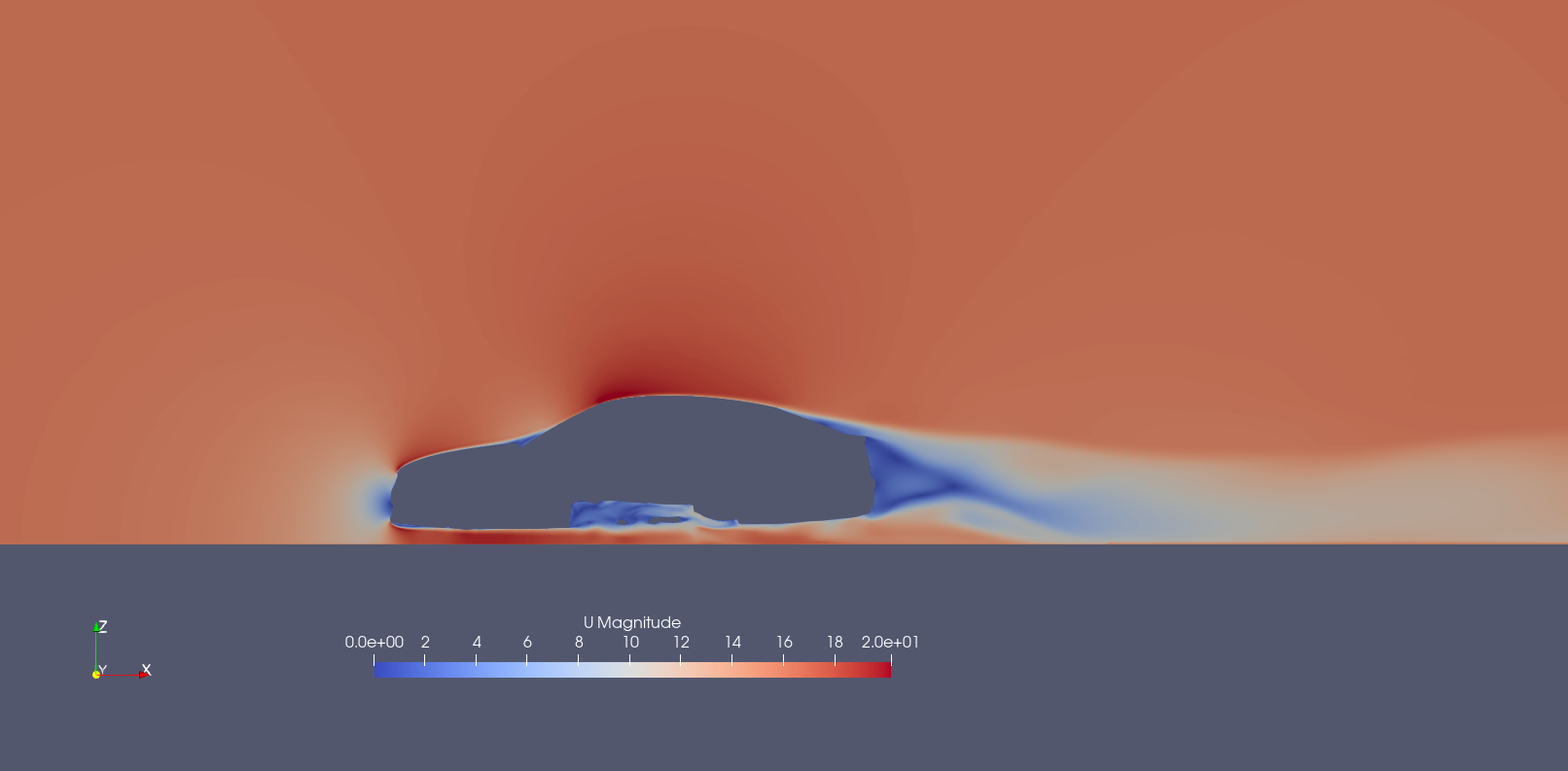}
		\caption{U-OpenFOAM-v2412}
		\label{fig:U-sub2}
	\end{subfigure}
	%% p
	% First Subfigure
	\begin{subfigure}[b]{0.48\textwidth}
		\centering
		\includegraphics[width=\textwidth]{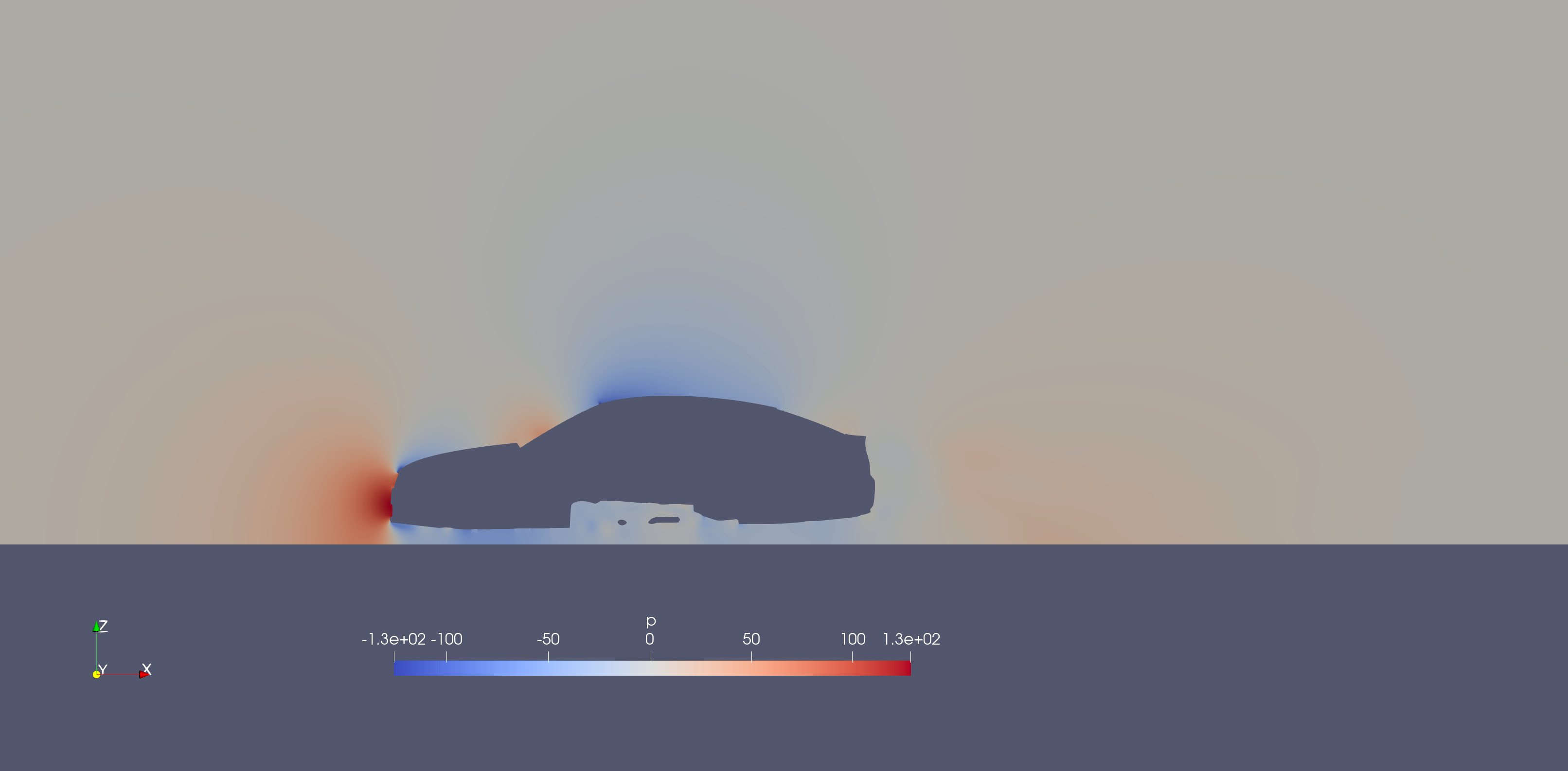}
		\caption{p-\myof{}}
		\label{fig:p-sub1}
	\end{subfigure}
	\hfill
	% Second Subfigure
	\begin{subfigure}[b]{0.48\textwidth}
		\centering
		\includegraphics[width=\textwidth]{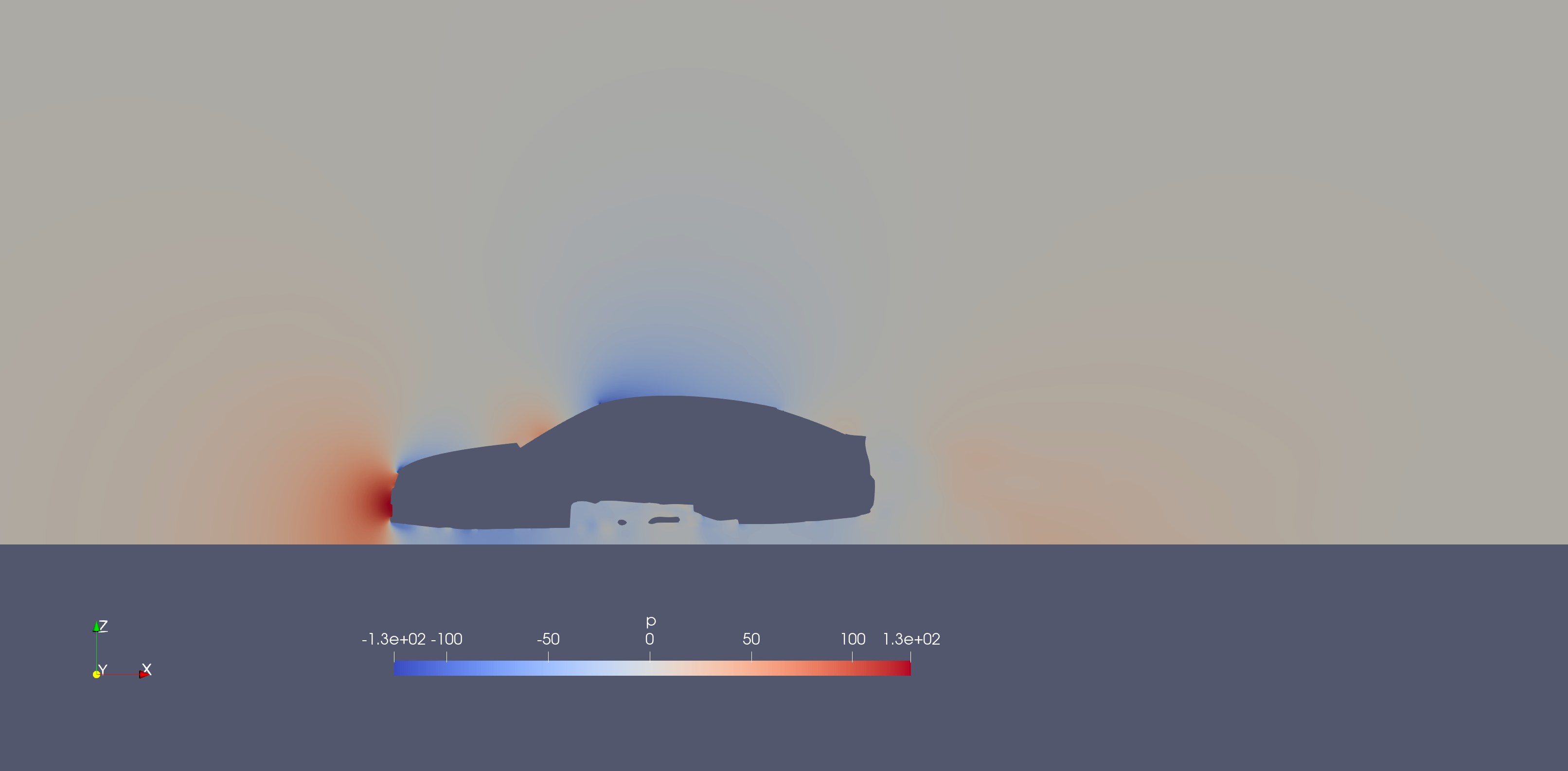}
		\caption{p-OpenFOAM-v2412}
		\label{fig:p-sub2}
	\end{subfigure}
	%% eddy
	% First Subfigure
	\begin{subfigure}[b]{0.48\textwidth}
		\centering
		\includegraphics[width=\textwidth]{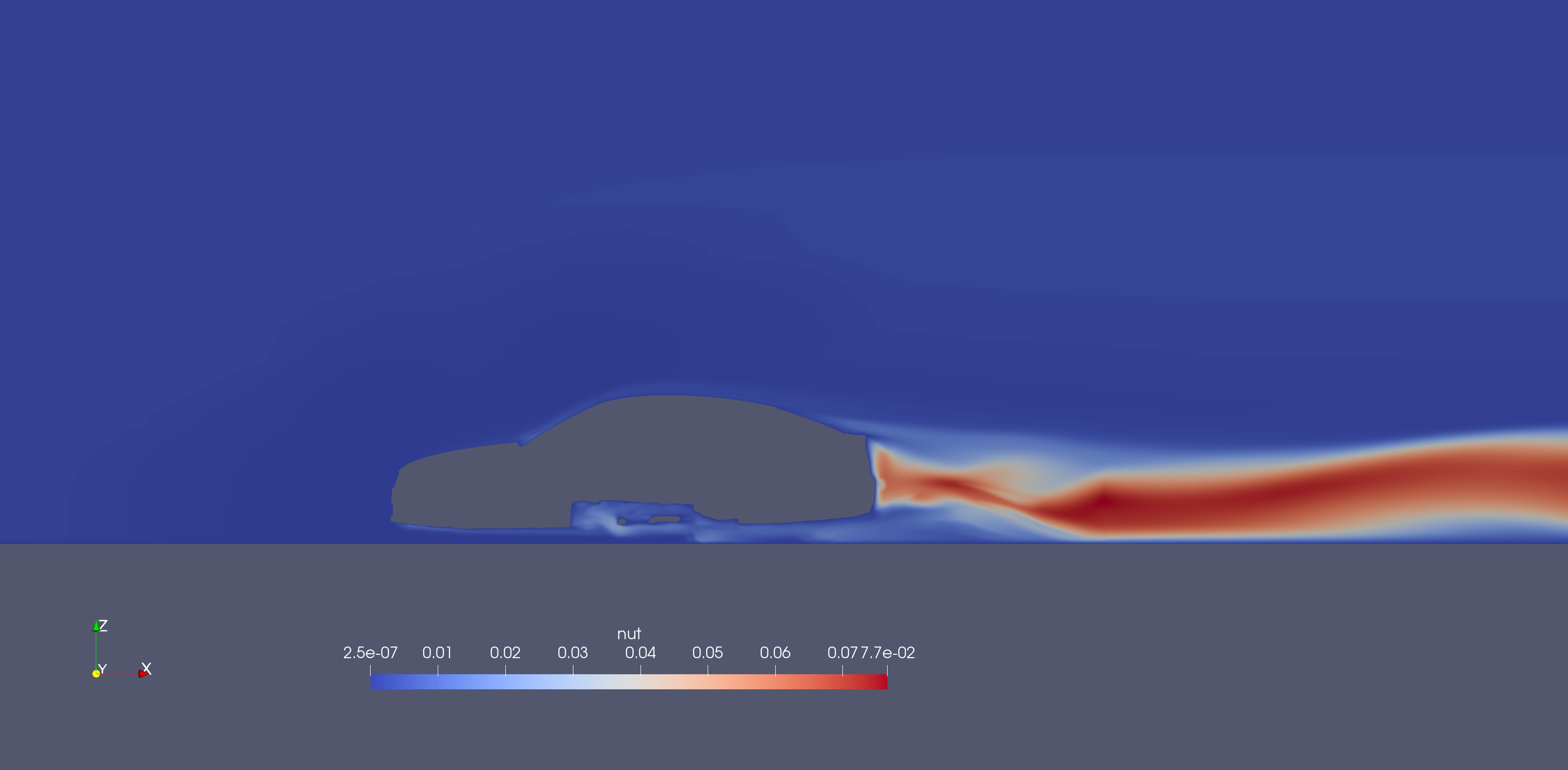}
		\caption{eddy viscosity-\myof{}}
		\label{fig:eddy-sub1}
	\end{subfigure}
	\hfill
	% Second Subfigure
	\begin{subfigure}[b]{0.48\textwidth}
		\centering
		\includegraphics[width=\textwidth]{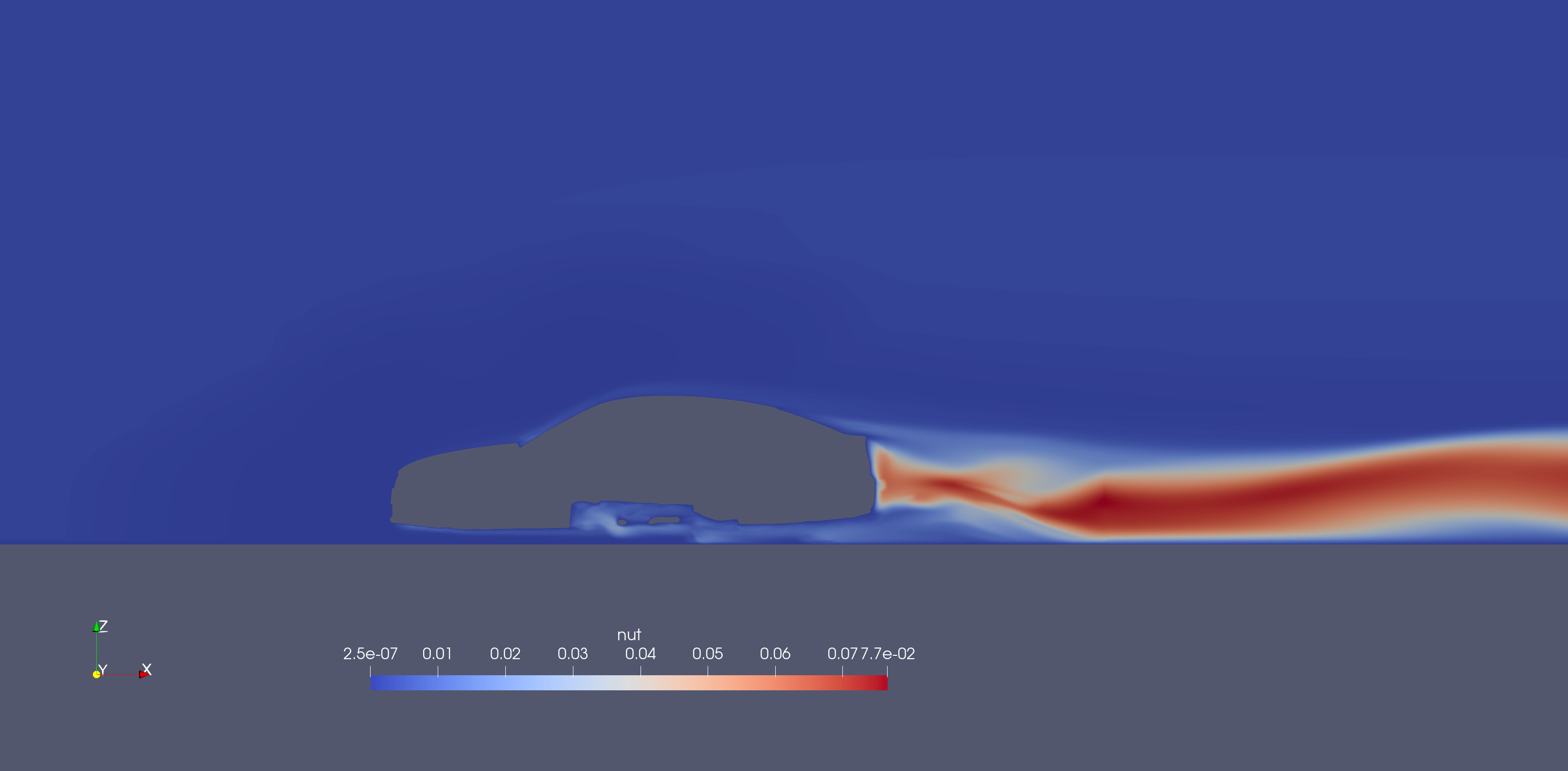}
		\caption{eddy viscosity-OpenFOAM-v2412}
		\label{fig:eddy-sub2}
	\end{subfigure}
	\caption{DrivAer22M. Comparison of velocity field (top), pressure field (middle) and eddy viscosity (bottom) at T=2100 between \myof{} and OpenFOAM-v2412 with PCG as pressure solver.}
	\label{fig:U_p_eddy_diff_field}
\end{figure}

A comparison between mean $C_d$ and $C_l$ obtained with different pressure solvers is presented in Table \ref{tab:mean_force_coeff_T_2000}, showing good agreement between \myof{} and OpenFOAM-v2412. Good agreement is also evident from the visual comparison of the flow fields. The velocity, pressure and eddy viscosity fields on the symmetry plane of the car after $T=2100$ time steps are shown in Figure \ref{fig:U_p_eddy_diff_field}.

%%%%%%%%%%%%%%%%%%%%%%%%%%%%%%%%%%%%%%%%%%%%%%%%%%%%%%%%%%
%%%%%%%%%%%%%%%%%%%%%%%%%%%%%%%%%%%%%%%%%%%%%%%%%%%%%%%%%%

%%%%%%%%%%%%%%%%%%%%%%%%%%%%%%%%%%%%%%%%%%%%%%%%%%%%%%%%%%
\begin{figure}[ht]
	\centering
	%% First subfigure
	\begin{subfigure}[b]{0.48\textwidth}
		\centering
		\includegraphics[width=\textwidth]{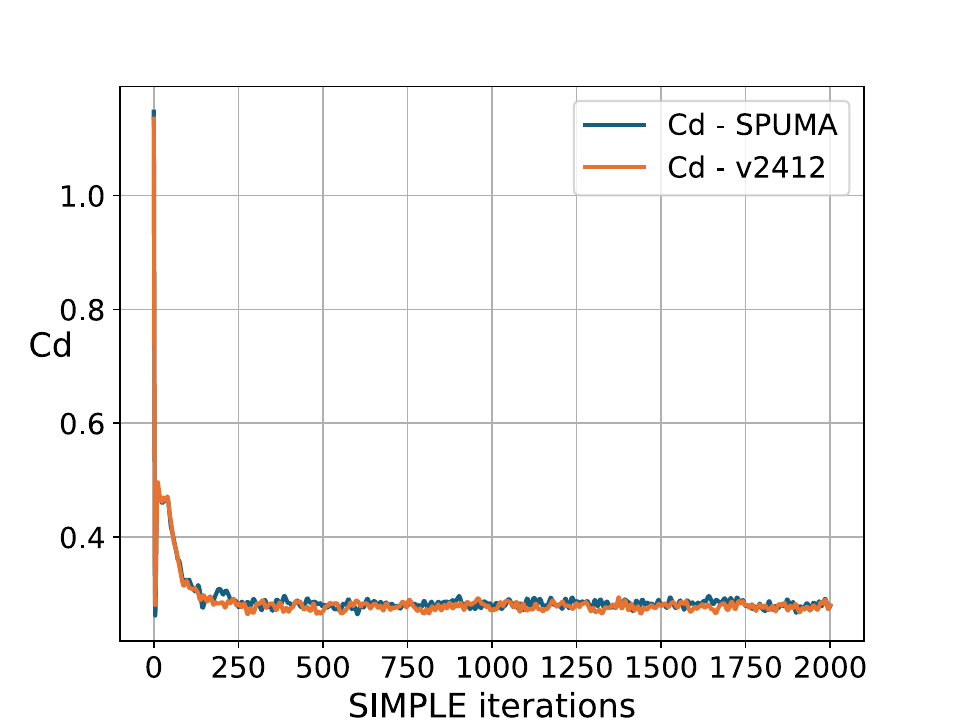} %% img/validation
		\caption{}
		\label{fig:cd-sub1}
	\end{subfigure}
	\hfill
	% Second Subfigure
	\begin{subfigure}[b]{0.48\textwidth}
		\centering
		\includegraphics[width=\textwidth]{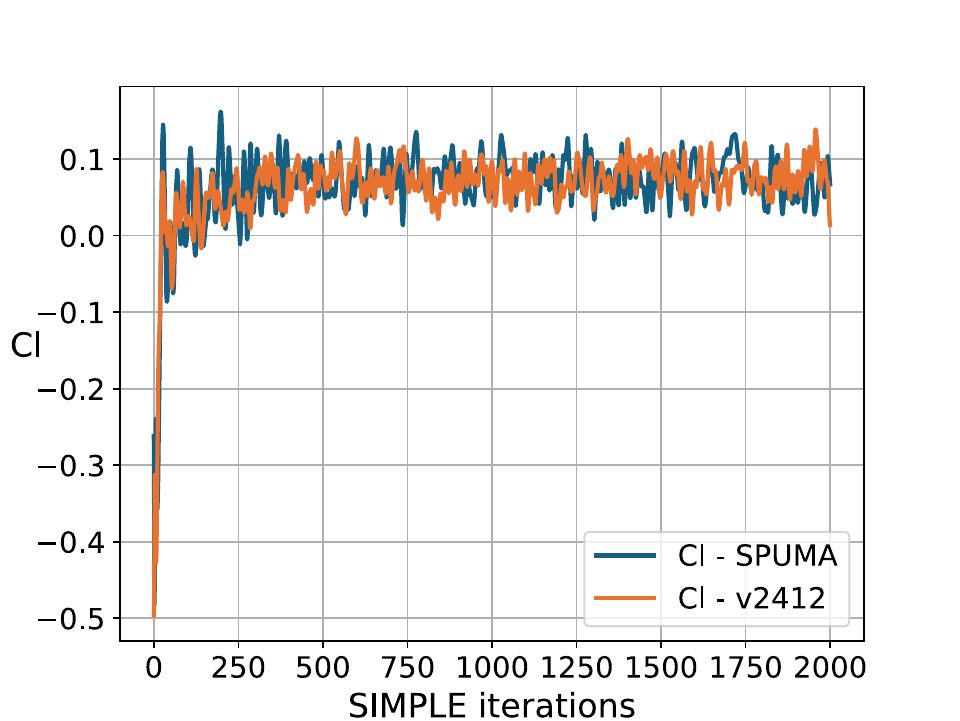}
		\caption{}
		\label{fig:cl-sub2}
	\end{subfigure}
	\caption{occDrivAer65M. Drag (a) and lift (b) coefficient comparison}
	\label{fig:cd_cl_comp_mg}
\end{figure}
%%%%%%%%%%%%%%%%%%%%%%%%%%%%%%%%%%%%%%%%%%%%%%%%%%%%%%%%%%

%%%%%%%%%%%%%%%%%%%%%%%%%%%%%%%%%%%%%%%%%%%%%%%%%%%%%%%%%%
\begin{table}[ht]
	\centering
	\begin{tabular}{c c c c c}
		\hline
		& \multicolumn{2}{c}{$C_d$} & \multicolumn{2}{c}{$C_l$}  \\
		& Leonardo & LUMI & Leonardo & LUMI \\
		\hline
		OpenFOAM-v2412 (GAMG) & 0.27847 & 0.28017 & 0.07793 & 0.07822 \\
		\myof{} (GAMG) & 0.28031 & 0.28167 & 0.07812 & 0.07881 \\
		\hline
	\end{tabular}
	\caption{occDrivAer65M. Mean drag and lift coefficients using different hardware. The average procedure is performed in the interval [$T_i=500, T_f=2000$]}
	\label{tab:mean_force_coeff}
\end{table}
%%%%%%%%%%%%%%%%%%%%%%%%%%%%%%%%%%%%%%%%%%%%%%%%%%%%%%%%%%

Another set of simulations was performed analyzing the RANS occDrivAer65M test case with OpenFOAM-v2412 (800 cores) and \myof{} (8 GPUs), running the case in parallel up to 2000 iterations. The last 1500 pseudo-time iteration window was used to evaluate the mean force coefficients. The drag and lift coefficients are plotted as a function of time in Figure \ref{fig:cd_cl_comp_mg}, while the average values are shown in Table \ref{tab:mean_force_coeff}. 
The relative difference is less than $1\%$ for both the drag and lift coefficients.
%, showing a higher sensitivity for the latter to the decomposition (800 vs 8 partitions). 
This small discrepancy is due to the absence of a stationary solution and a strong dependence of the solution on the error propagation. Throughout hundreds of iterations, the time history of mean force coefficients diverges, showing a chaotic nature, see Figure \ref{fig:cd_cl_comp_mg}.\\ Since floating point algebra is not associative and distributive up to machine precision, results can be slightly influenced by the order in which operations are performed. These small perturbations are amplified in the wake and in the underside regions of the car, affecting the force coefficients. Any change to the case setup, such as a different domain partitioning or compiler optimization flags, will produce different results to some extent. On GPUs, the order in which parallel operations are performed is random, further reducing reproducibility.

\subsection{Profiling}
\label{subsec::profiling_results}
A profiling of the application's low-level routines and memory utilization is carried out using the CUDA backend, both in serial and in parallel (2 GPUs).
In Table \ref{tab:time_breakdown_cuda1} we report the percentage time breakdown of the CUDA kernels and memory operations relative to the total CUDA context time spent in five SIMPLE iterations of the DrivAer22M test case (from iteration 2 to iteration 6, ignoring the first iteration). Approximately 83\% of the time is spent in kernel execution (i.e. \textit{reduction} and \textit{parallel\_for}), while the remaining part is left to memory operations (i.e. \textit{memset} and host-device \textit{memcopy}). Additional time percentages are provided for \textit{memcopy} operations and \textit{parallel\_for} kernels. For the latter, all computationally intensive kernels---primarily Sparse Matrix-Vector Multiplications (SpMVM) and gradient evaluations---are included, except those contributing less than 0.1\% each. Among \textit{memcopy} operations, the most significant cost is the device-to-device transfer, which accounts for approximately 53\% of the total time.  

%%%%%%%%%%%%%%%%%%%%%%%%%%%%%%%%%%%%%%%%%%%%%%%%%%%%%%%%%%%%%%%%%%%%%%
\begin{table}[p]
	\centering
	\resizebox{0.55\textwidth}{!}
	{
		\begin{tabular}{l c l c}
			\hline
			Task & Time \% & Item & Time \% \\
			\hline
			\multirow{6}{*}{parallel for} & \multirow{6}{*}{79.27}   & SpMVM (Amul + Tmul) & 21.79 \\
			&                          & gaussGrad::gradf               & 15.06  \\
			&                          & cellLimitedGrad     & 13.94  \\        
			&                          & aDILUPreconditioner & 9.07  \\
			&                          & multiply            & 5.16  \\
			&                          & surfaceInterpolationScheme::dotInterpolate   & 4.05  \\ 
			&                          & -= & 3.26  \\
			&                          & surfaceIntegrate    & 3.22  \\
			&                          & /= & 2.84  \\                          
			&                          & PBiCG::solve & 2.13  \\ 
			&                          & += & 2.10  \\
			&                          & RichardsonSmoother::smooth  & 1.88  \\
			&                          & subtract & 1.23  \\
			&                          & lduMatrix::sumA & 1.06  \\
			&                          & lduMatrix::negSumDiag & 1.04  \\
			&                          & lduMatrix::H + lduMatrix::H1 & 0.83  \\
			&                          & fvPatch::patchInternalField & 0.82  \\              
			&                          & add & 0.80  \\
			&                          & Field::negate & 0.66  \\
			&                          & GAMGSolver::scale & 0.64  \\
			&                          & divide & 0.55  \\
			&                          & GAMGAgglomeration::restrictField & 0.55  \\   
			&                          & lduMatrix::sumMagOffDiag & 0.48  \\
			&                          & GAMGSolver::intefaceInternalField & 0.46  \\
			&                          & *= & 0.42  \\
			&                          & component & 0.41  \\
			&                          & Field::replace & 0.40  \\
			&                          & T & 0.40  \\
			&                          & dev2 & 0.39  \\
			&                          & linearUpwind & 0.38  \\
			&                          & lduInterfaceField::addToInternalField & 0.32  \\
			&                          & GAMGSolver::prolongField & 0.32  \\
			&                          & GAMGSolver::agglomerateMatrix & 0.30  \\
			&                          & mag & 0.27  \\
			&                          & max & 0.25  \\
			&                          & symm & 0.24  \\
			&                          & devTwoSymm & 0.23  \\
			&                          & min & 0.23  \\
			&                          & dot & 0.22  \\
			&                          & GAMGSolver::Vcycle & 0.2  \\
			&                          & pos0 & 0.18  \\
			&                          & dotdot & 0.17  \\
			&                          & outer & 0.14  \\
			&                          & fvMatrix::setValuesFromList & 0.14  \\
			&                          & sqr & 0.14  \\
			&                          & sqrt & 0.11  \\
			&                          & snGradScheme::snGrad & 0.10  \\
			&                          & lduMatrix::faceH & 0.10  \\
			&                          & other ($<$ 0.1\% each)  & 0.27  \\
			reduction                     &  4.18                    &                     &          \\
			custom memset                 &  9.88                    &                     &          \\
			\multirow{6}{*}{memory ops}   &  \multirow{6}{*}{6.67} & D2D                 & 52.70 \\
			&                          & H2D                 & 18.62 \\
			&                          & D2H                 & 15.44 \\
			&                          & memset              & 11 \\
			&                          & unified H2D         & 1.24 \\
			&                          & unified D2H         & 1 \\
			\hline
		\end{tabular}
	}
	\caption{Percentage time breakdown of the CUDA kernel computation and memory operations relative to the total CUDA context time. This run is performed on Leonardo, using the DrivAer22M test case for five iterations ($T_f - T_i$, $T_f=6$, $T_i=1$) with 2 GPUs, GAMG as pressure solver (coarsestSolver=DIC+PCG) and the \textit{fixedSize} memory pool.}
	\label{tab:time_breakdown_cuda1}
\end{table}
%%%%%%%%%%%%%%%%%%%%%%%%%%%%%%%%%%%%%%%%%%%%%%%%%%%%%%%%%%%%%%%

%%%%%%%%%%%%%%%%%%%%%%%%%%%%%%%%%%%%%%%%%%%%%%%%%%%%%%%%%%%%%%%

\begin{table}[H]
	\centering
	\resizebox{1.\textwidth}{!}
	{
		\begin{tabular}{|l|c|c|c|c|c|c|c|c|c|}
			\hline
			\textbf{Operation} & \textbf{Pool Type} &  \textbf{Time [\%] } & \textbf{Total Time} [ms] & \textbf{Count} & \textbf{Average} [$\mu$s] & \textbf{Median} [$\mu$s] & \textbf{Total size} & \textbf{Average} & \textbf{Median} \\
			\hline
			memset & dummy & 28.0\% & 3757 & 6526 & 575.652 & 1.729 & 66.34 GiB & 10.41 MiB & 232 B \\
			unified H2D & dummy & 0.0 & 3.044 & 1207 & 2.521 & 2.272 & 12.13 MiB & 10.29 KiB & 4.00 KiB \\
			unified D2H & dummy & 0.0 & 2.508 & 1205 & 2.081 & 1.728 & 12.13 MiB & 10,31 KiB & 4,00 KiB \\
			H2D & dummy & 0.0 & 85.081 & 4244 & 20.047 & 4.640 & 49.95 KiB	& 12 B & 8 B \\
			D2H & dummy & 0.0 & 10.907 & 3156 & 3.455 & 3.264 & 24.66 KiB & 8 B & 8 B \\
			D2D & dummy & 71.0 & 9455 & 3602 & 2625 & 185.807 & 230.75 GiB & 65.60 MiB & 2.72 MiB \\
			\hline
			memset & fixedSize & 11.0 & 56.411 & 6526 & 8.644 & 1.696 & 66.34 GiB & 10.41 MiB & 232 B \\
			unified H2D & fixedSize & 0.0 & 2.764 & 1101 & 2.510 & 2.335 & 10.84 MiB & 10.08 KiB & 4.00 KiB \\
			unified D2H & fixedSize & 0.0 & 2.513 & 1300 & 1.932 & 1.727 & 10.84 MiB & 8.54 KiB & 4.00 KiB \\
			H2D & fixedSize & 8.0 & 40.846 & 4244 & 9.624 & 4.032 & 49.95 KiB & 12 B & 8 B \\
			D2H & fixedSize & 2.0 & 10.603 & 3156 & 3.359 & 3.264 & 24.66 KiB & 8 B & 8 B \\
			D2D & fixedSize & 77.0 & 387.542 & 3602 & 107.590 & 8.704 & 230.75 GiB & 65.60 MiB & 2.72 MiB \\
			\hline
		\end{tabular}
	}
	\caption{Memory transfer statistics for five iterations ($T_f - T_i$, $T_f=6$, $T_i=1$) of \textit{simpleFoam}, running the DrivAer22M test case with \myof{} using 1 GPU. Both \textit{dummy} and \textit{fixedSize} memory pools are tested.}
	\label{tab:mem_1gpu}
\end{table}

\begin{table}[H]
	\centering
	\resizebox{1.\textwidth}{!}
	{
		\begin{tabular}{|l|c|c|c|c|c|c|c|c|c|}
			\hline
			\textbf{API} & \textbf{Pool Type} &  \textbf{Time [\%] } & \textbf{Total Time} [ms] & \textbf{Count} & \textbf{Average} [$\mu$s] & \textbf{Median} [$\mu$s] & \textbf{Min} [$\mu$s] & \textbf{Max} [$\mu$s] \\
			\hline
			cudaDeviceSynchronize & dummy & 72.8\% & 54542 & 53259 & 1024   & 8.218    & 1.584 & 326289 \\
			cudaMemcpy            & dummy & 12.7\% & 9536 & 6758 & 1411  & 14.144 & 9.888 & 31060 \\
			cudaFree            & dummy & 8.8\% & 6574 & 27192 & 241.761  & 9.962 & 5.464 & 13300 \\
			cudaMemset            & dummy & 5.1\% & 3812 & 6526 & 584.154  & 10.041 & 8.006 & 31982 \\
			cudaMallocManaged            & dummy & 0.3\% & 233.885 & 27192 & 8.601  & 6.529 & 1.499 & 367.717 \\
			cudaLaunchKernel            & dummy & 0.2\% & 164.042 & 53259 & 3.080  & 2.756 & 2.266 & 35.659 \\
			cudaMemcpyAsync            & dummy & 0.0\% & 28.521 & 4244 & 6.720  & 5.876 & 4.391 & 24.201 \\
			\hline
			cudaDeviceSynchronize & fixedSize & 92.0\% & 8403 & 53259 & 157.784   & 7.788    & 1.566 & 112744 \\
			cudaMemcpy            & fixedSize & 5.0\% & 460.657 & 6758 & 68.164  & 13.144 & 9.437 & 859.092 \\
			cudaLaunchKernel      & fixedSize & 1.6\% & 144.224 & 53259 & 2.708   & 2.550   & 2.278 & 241.593 \\
			cudaMemset            & fixedSize & 1.2\%	& 106.062 & 6526 & 16.252 & 9.304  & 8.094 & 383.221 \\	
			cudaMemcpyAsync       & fixedSize & 0.3\%	& 24.014 & 4244	& 5.658   & 5.203  & 4.179 & 41.653 \\	
			\hline
		\end{tabular}
	}
	\caption{CUDA API statistics for five iterations ($T_f - T_i$, $T_f=6$, $T_i=1$) of \textit{simpleFoam}, running the DrivAer22M test case with \myof{} using 1 GPU. Both  \textit{dummy} and \textit{fixedSize} memory pools are tested. Since every kernel invocation  in SPUMA is followed by a call to \textit{cudaDeviceSynchronize}, this API collects the time spent in the kernel execution and the overhead associated to synchronization.}
	\label{tab:api_1gpu}
\end{table}

Memory operations are further analyzed from a statistical point of view in Tables \ref{tab:mem_1gpu} and \ref{tab:api_1gpu}, using one GPU and the \textit{fixedSize} and \textit{dummy} memory pools. The analysis shows the dramatic impact of adopting the memory pool in reducting the time-to-solution: the cost associated with \textit{memsets}, device-to-device copies and kernel execution (wrapped into \textit{cudaDeviceSynchronize}) is reduced by one order of magnitude or more, due to the less fragmented memory strongly reducing non-contiguous memory accesses.
	The overhead of memory copies from host to device and vice versa is only slightly reduced. The memory pool also reduces the cost associated with allocation and deallocation of objects by removing tens of thousands of invocations to \textit{cudaMallocManaged} and \textit{cudaFree}, as shown in Table \ref{tab:api_1gpu}.

%%%%%%%%%%%%%%%%%%%%%%%%%%%%%%%%%%%%%%%%%%%%%%%%%%%%%%%%%%%%%%%%%%%%%%

\begin{table}[ht]
	\centering
	\begin{tabular}{c c c c c}
		\hline
		Case & no pool & dummy & fixedSize & Umpire  \\
		\hline
		DrivAer22M-200 Cores & 154.43 s & 145.18 s & 144.64 s & 154.37 s\\
		DrivAer22M-400 Cores & 64.29 s  & 62.40 s & 61.25 s   & 63.97 s\\
		DrivAer22M-1 GPU & - & 822.68 s & 141.40 s & 140.63 s \\
		DrivAer22M-2 GPU & - & 431.81 s & 77.14 s & 77.55 s \\
		\hline
	\end{tabular}
	\caption{Execution time spent in 100 iterations starting from T=10 with 3 different memory pools (implemented in \myof{}) and without a memory pool (OpenFOAM-v2412).}
	\label{tab:mem_pool_impact}
\end{table}

%%%%%%%%%%%%%%%%%%%%%%%%%%%%%%%%%%%%%%%%%%%%%%%%%%%%%%%%%%%%%%%%%%%%%
%%%%%%%%%%%%%%%%%%%%%%%%%%%%%%%%%%%%%%%%%%%%%%%%%%%%%%%%%%%%%%%

\begin{figure}[h]
	\centering
	\includegraphics[width=0.8\linewidth]{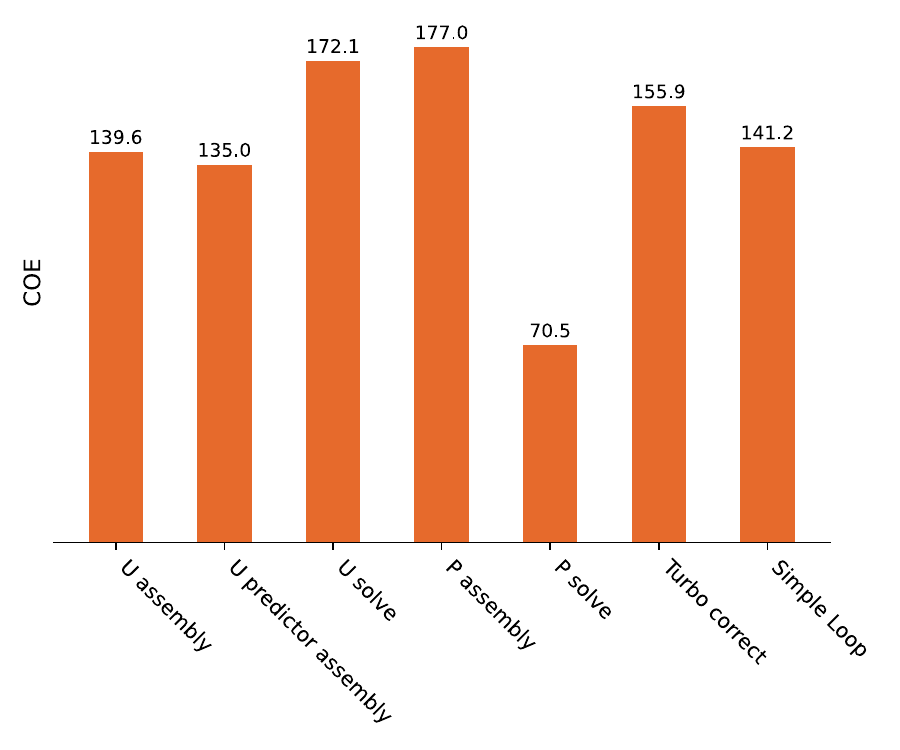}
	\caption{Coefficient of equivalence (COE) breakdown for a \textit{simpleFoam} run using \myof{} on 2 GPUs and OpenFOAM-v2412 on 400 CPU cores. The GAMG solver is used for the pressure equation, with the PCG solver and DIC preconditioner for the coarsest level. The simulation is performed on the Leonardo HPC cluster using the DrivAer22M test case.}
	\label{fig:section_speedup}
\end{figure}

%%%%%%%%%%%%%%%%%%%%%%%%%%%%%%%%%%%%%%%%%%%%%%%%%%%%%%%%%%%%%%%%%%%%%%
The advantage with the memory pool is also highlighted by the results of Table \ref{tab:mem_pool_impact}, which reports the speed-up of SPUMA compared to OpenFOAM-v2412 using three different memory pool strategies. The speed-up is not limited to GPUs, where the performance gain is remarkable (about 4X) using the \textit{fixedSize} or the \textit{Umpire} memory pool, but also CPU simulations show a reduction in the time-to-solution. A third-party library like Umpire, implementing a more sophisticated dynamic pool, is not providing any performance improvement compared to the \textit{fixedSize} memory pool developed by the authors.

%equivalent cores/gpu speedup metric 
To compare performances between GPU and CPU runs, we used a Coefficient Of Equivalence (COE)  defined as the ratio between the time of a CPU run multiplied by the number of cores and the time of a GPU run multiplied by the number of GPUs, as shown in equation \ref{eq:coe}. 
	\begin{equation}
		COE = \frac{T_{cpu} * N_{cores}}{T_{gpu} * N_{gpus}}
		\label{eq:coe}
	\end{equation}
	Figure \ref{fig:section_speedup} shows values of the COE obtained over 100 SIMPLE iterations starting from $T=1500$, for six different main parts of the SIMPLE algorithm (e.g. assembly of the momentum equations). The plot is not uniform: a peak of 180 is achieved in the solution of the momentum equations and in the assembly process of the pressure linear system while a lower value of about 140 is obtained in the rhs evaluation of the momentum equations characterized by the gradient computation, a kernel containing several atomic operations. However, the lowest value, less than 100, is achieved in the pressure solution using the GAMG solver.

Effective GPU parallelization of the GAMG solver is not a trivial task, it depends on the coarsening algorithm and on the type of smoother. In this work, we implemented two parallel smoothers running completely on GPUs, namely Richardson (or weighted Jacobi) and two-stage Gauss-Seidel, as described in \cite{bergervergiat2021}, but the coarsening algorithm has not been modified. The effect of changing the smoother only at the coarse level or in all levels of the hierarchy is reported in Table \ref{tab:parallel_smoother}. The fastest solution is obtained using Richardson in all levels and diagonal at the coarsest one. Using a more effective CPU preconditioner as DIC, which involves data transfer, does not reduce the walltime. An extensive analysis on this topic is left to a future work. 

%%%%%%%%%%%%%%%%%%%%%%%%%%%%%%%%%%%%%%%%%%%%%%%%%%%%%%%%%%%%%%%%%%%%%%

\begin{table}[ht]
	\begin{tabular}{cc|cc|}
		\cline{3-4}
		&                            & \multicolumn{2}{c|}{Coarse level preconditioner} \\ \cline{3-4} 
		&                            & \multicolumn{1}{c|}{DIC}          & diagonal     \\ \hline
		\multicolumn{1}{|c|}{\multirow{2}{*}{Smoother}} & Richardson (damped Jacobi) & \multicolumn{1}{c|}{122.59 s}     & 115.76 s     \\ \cline{2-4} 
		\multicolumn{1}{|c|}{}                          & two-stage-Gauss-Seidel     & \multicolumn{1}{c|}{132.02 s}     & 123.9 s      \\ \hline
	\end{tabular}
	\caption{Walltimes for 100 SIMPLE iterations of the DrivAer22M test case, using different smoothers and coarse level preconditioners in the GAMG solver for the pressure equation.}
	\label{tab:parallel_smoother}
\end{table}

%%%%%%%%%%%%%%%%%%%%%%%%%%%%%%%%%%%%%%%%%%%%%%%%%%%%%%%%%%%%%%%%%%%%%

\subsection{Scalability test}
To assess parallel performance, both weak and strong scalability tests were performed on the occDrivAer benchmark \cite{occ_drivaer} with OpenFOAM's hierarchical decomposition method.
Strong scaling tests were performed on the smallest mesh, the 65M case, while weak scaling tests were performed for all three meshes reported in Table \ref{tab:drivaer-meshes}. Because the maximum number of cores per node differs between the CPU partitions of Leonardo and LUMI (see Tables \ref{tab:leonardo_hardware} and \ref{tab:lumi_hardware}), all simulations were run with 100 cores per node on both clusters. This choice is slightly below the maximum core count for both clusters but remains close to it. Under-subscribing these nodes---characterized by large caches and high core counts---is common practice, as it helps absorb background system tasks and housekeeping without degrading job performance. Thus, using one hundred cores represents a good compromise between consistency across clusters, smooth execution, and divisibility for obtaining a regular domain decomposition (e.g. hierarchical).

\begin{figure}[ht]
	\centering
	%% First subfigure
	\begin{subfigure}[b]{0.96\textwidth}
		\centering
		\includegraphics[width=\textwidth]{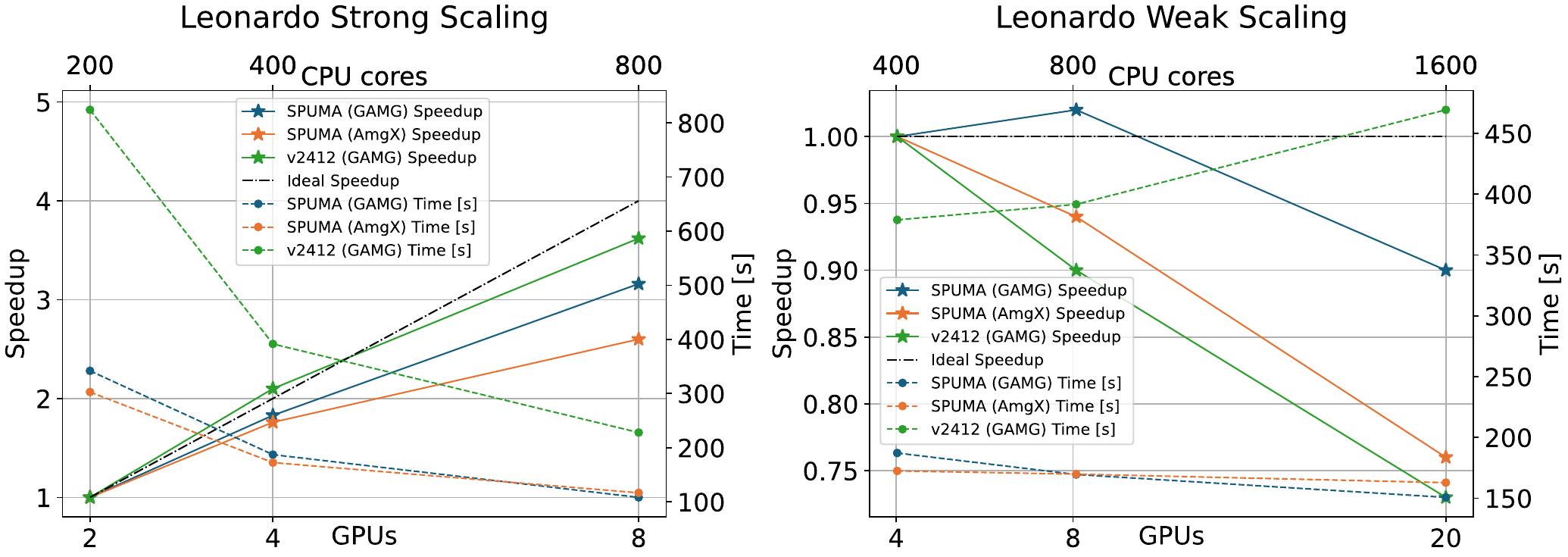} %% scaling
		\caption{}
		\label{fig:strong-scaling-sub1}
	\end{subfigure}
	\hfill
	% Second Subfigure
	\begin{subfigure}[b]{0.96\textwidth}
		\centering
		\includegraphics[width=\textwidth]{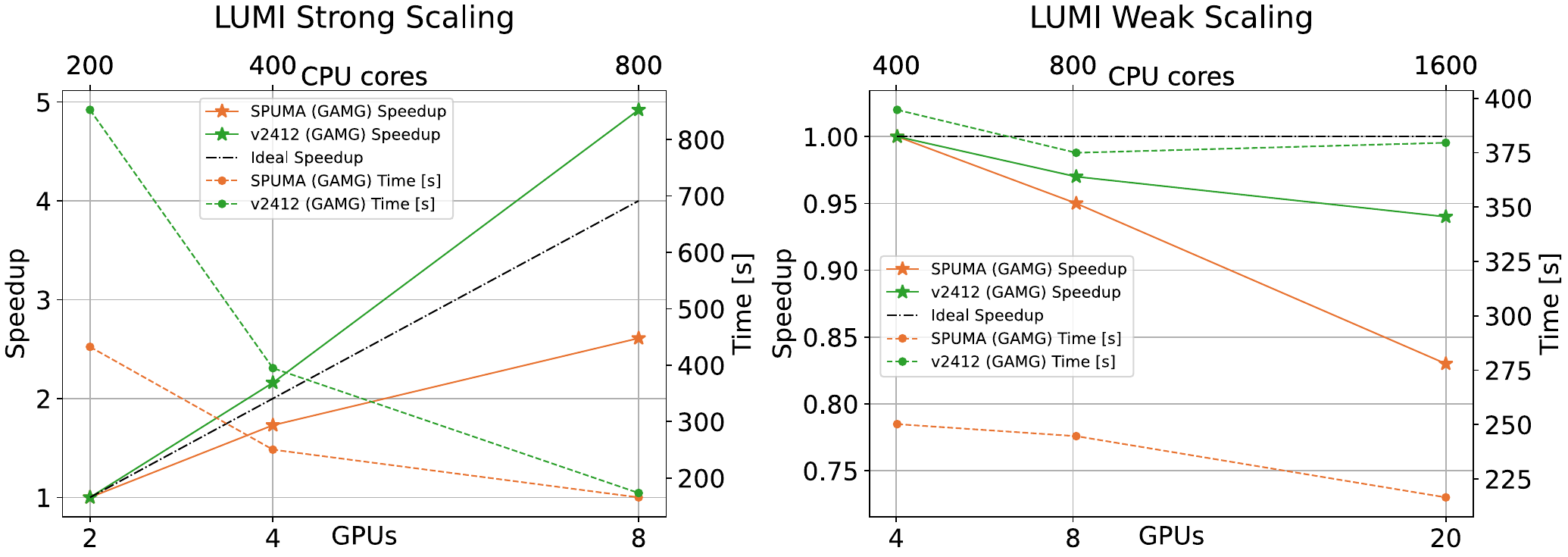} %% scaling
		\caption{}
		\label{fig:weak-scaling-sub2}
	\end{subfigure} \\
	\caption{Strong and weak scaling performances of \myof{}, with AMGX (Leonardo only) and GAMG solvers for pressure, compared with OpenFOAM-v2412 on Leonardo (a) and LUMI (b). Speedup is measured relative to the smallest decomposition. The time reported is that required by 100 SIMPLE iterations using the occDrivAer test case.}
	\label{fig:weak_strong_scaling_drivaer_LEO_LUMI}
\end{figure}

\begin{figure}[h!]
	\centering
	
	\begin{subfigure}[]{0.9\textwidth}
		\centering
		\includegraphics[width=\textwidth]{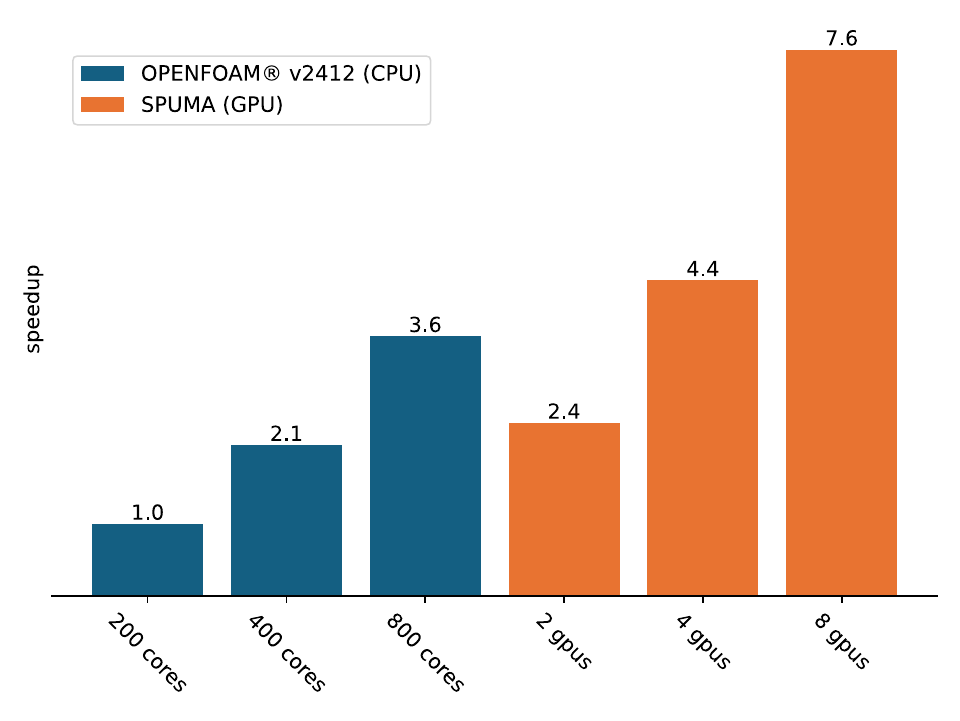}
		\caption{}
	\end{subfigure}
	
	\hfill
	
	\begin{subfigure}[]{\textwidth}
		\centering
		\includegraphics[width=\textwidth]{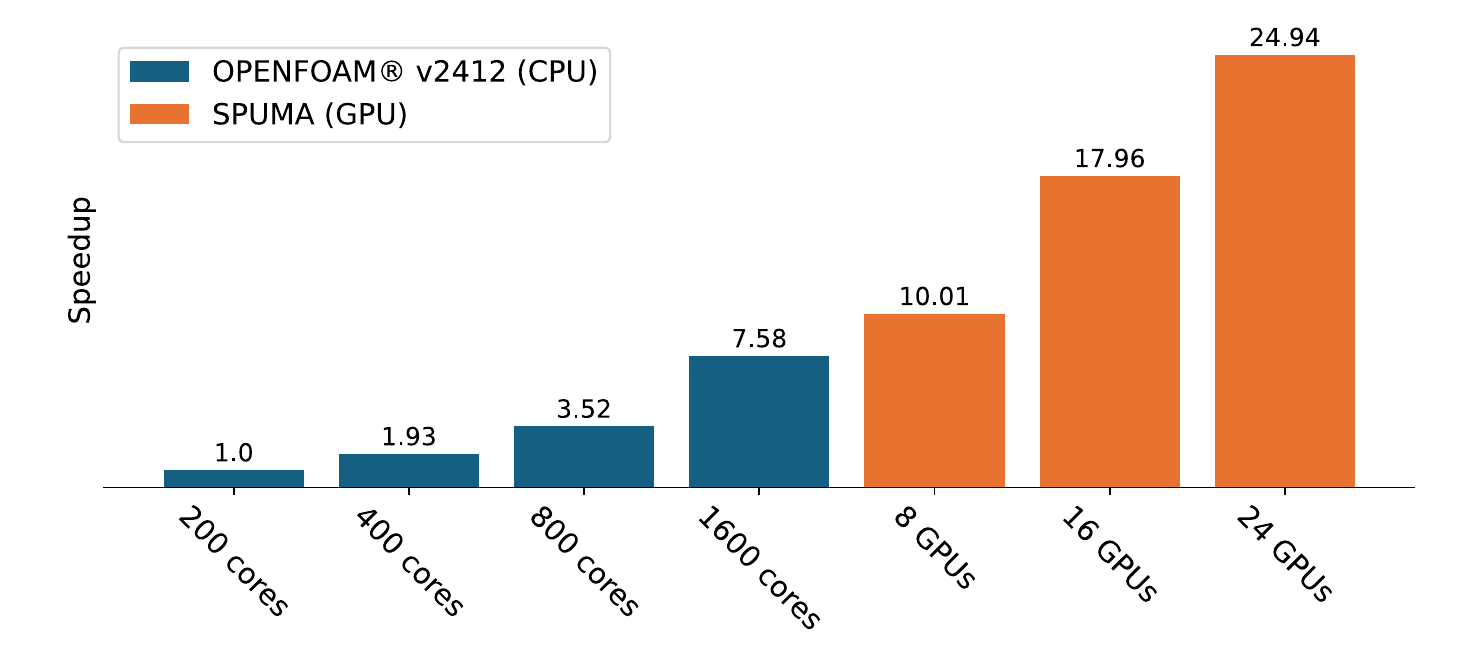}
		\caption{}
	\end{subfigure}
	
	\caption
	{Strong scaling results of the occDrivAer65M test case (a) and of the occDrivAer235M test case (b) : speedups relative to the 200 cores CPU run; \myof{} use AMGX as pressure solver.}
	\label{fig:speedup_occDrivAer}
\end{figure}

Figure \ref{fig:weak_strong_scaling_drivaer_LEO_LUMI} shows weak and strong scaling performances (both speed-up and computational time) of the whole solver, while tables \ref{tab:leo_weak_scalability}, \ref{tab:leo_strong_scalability}, \ref{tab:lumi_weak_scalability} and \ref{tab:lumi_strong_scalability} offer a more detailed breakdown of the time spent in each section of the solver. Figure \ref{fig:speedup_occDrivAer} shows the speed-up of the occDrivAer65M and occDrivAer235M cases relative to a base case run with OpenFOAM-v2412 on 200 cores (two nodes) of Leonardo's data centric partition.

Regarding the strong scalability, the curve is the same for both Leonardo and LUMI using the GAMG solver. Efficiency is still good at four GPUs, but it starts to drop at 8 GPUs reaching a value of about 65\%. Switching to AmgX as pressure solver (available on Leonardo only), the efficiency increases up to 80\% at 8 GPUs due to the aggressive coarsening feature not available in the built-in GAMG solver. 

Since AmgX is a pure algebraic multigrid solver and the matrix coefficients are changing from one time step to the other in the SIMPLE loop, the coarse grids of the multigrid hierarchy cannot be cached (as GAMG does) and need to be recomputed every time. This justifies the slower time of the pressure solver at 2 GPUs with respect to GAMG (2.5 times slower). However, the better scalability properties of AmgX allow for a faster wall time at 8 GPUs, demonstrating that GAMG still has unrealized potential. The reason for this behavior is that the GAMG solver, with default settings, generates numerous multigrid levels, many of which contain too few cells to effectively utilize the GPU and to overlap communication and computation. By contrast, AmgX's aggressive coarsening generates far fewer levels and automatically switches to CPU when the number of cells becomes too low. Improving the scalability of the GAMG solver and identifying optimal default settings are left to a future work.

The performances of OpenFOAM-v2412 on CPU, computed to get a reference for GPU simulations, are in line with the results shown in \cite{Galeazzo2024}: using few cores (i.e. 400) OpenFOAM-v2412 is slightly slower on LUMI with respect to Leonardo but it scales better and superlinearly, confirming that AMD CPUs behave better when the load per core is lower.

%% --- Leonardo Weak scaling ---%%
\begin{table}[h!]
	\centering
	\resizebox{1.\textwidth}{!}
	{
		\begin{tabular}{|l|c|c|c|c|c|c|c|c|c|}
			\hline
			& \multicolumn{3}{|c|}{\begin{tabular}{@{}c@{}}\textbf{SPUMA} \\ \textbf{GAMG} \end{tabular}} & \multicolumn{3}{|c|}{\begin{tabular}{@{}c@{}}\textbf{SPUMA} \\ \textbf{AmgX} \end{tabular}} & \multicolumn{3}{|c|}{\textbf{OpenFOAM-v2412}} \\
			\hline
			& \multicolumn{3}{|c|}{\textbf{\# GPUs}}  & \multicolumn{3}{|c|}{\textbf{\# GPUs}} & \multicolumn{3}{|c|}{\textbf{\# CPUs}}  \\
			\hline
			& \textbf{4} & \textbf{8} & \textbf{20}  & \textbf{4} & \textbf{8} & \textbf{20} & \textbf{400} & \textbf{800} & \textbf{1600} \\
			\hline
			\cellcolor[HTML]{DADADA} \textbf{\# cells} & 65.3 M & 121 M & 236 M & 65.3 M & 121 M & 236 M & 65.3 M & 121 M & 236 M \\ 
			\cellcolor[HTML]{DADADA} \textbf{\# cells/process} & 16.3 M & 15.1 M & 11.8 M & 16.3 M & 15.1 M & 11.8 M & 163 k & 151 k & 147 k \\
			\hline \rowcolor[HTML]{DADADA}
			\textbf{ T110-T10 [s]} & \textbf{172.41} & \textbf{169.79} &\textbf{ 162.79} & \textbf{187.13 }& \textbf{169.47} & \textbf{150.65} & \textbf{378.85} & \textbf{391.6} & \textbf{469.31} \\
			\hline
			\cellcolor[HTML]{DADADA} \textbf{SIMPLE loop [s]} & 1.748 & 1.735 & 1.696 & 1.894 & 1.721 & 1.530 & 3.844 & 3.994 & 4.792 \\
			\cellcolor[HTML]{DADADA} \textbf{U assembly [s]} & 0.377 & 0.339 & 0.269 & 0.375 & 0.337 & 0.269 & 0.460 & 0.413 & 0.439 \\
			\cellcolor[HTML]{DADADA} \textbf{U predictor [s]} & 0.022 & 0.019 & 0.015 & 0.022 & 0.019 & 0.015 & 0.034 & 0.029 & 0.023 \\
			\cellcolor[HTML]{DADADA} \textbf{U solve [s]} & 0.564 & 0.565 & 0.463 & 0.547 & 0.519 & 0.458 & 1.910 & 1.990 & 2.106 \\
			\cellcolor[HTML]{DADADA} \textbf{P assembly [s]} & 0.040 & 0.034 & 0.030 & 0.039 & 0.033 & 0.030 & 0.062 & 0.057 & 0.056 \\
			\cellcolor[HTML]{DADADA} \textbf{P solve [s]} & 0.124 & 0.253 & 0.475 & 0.304 & 0.302 & 0.317 & 0.355 & 0.527 & 1.148 \\
			\cellcolor[HTML]{DADADA} \textbf{Turbulence [s]} & 0.491 & 0.413 & 0.349 & 0.482 & 0.404 & 0.350 & 0.816 & 0.777 & 0.810 \\
			\hline
		\end{tabular}
	}
	\caption{Detailed section breakdown of weak scaling performances of \myof{}, with AMGX and GAMG solver for pressure, compared with OpenFOAM-v2412, on Leonardo HPC system using the occDrivAer test case.}
	\label{tab:leo_weak_scalability}
\end{table}

%% --- Leonardo strong scaling ---%%
\begin{table}[h!]
	\centering
	\resizebox{1.\textwidth}{!}
	{
		\begin{tabular}{|l|c|c|c|c|c|c|c|c|c|}
			\hline
			& \multicolumn{3}{|c|}{\begin{tabular}{@{}c@{}}\textbf{SPUMA} \\ \textbf{GAMG} \end{tabular}} & \multicolumn{3}{|c|}{\begin{tabular}{@{}c@{}}\textbf{SPUMA} \\ \textbf{AmgX} \end{tabular}} & \multicolumn{3}{|c|}{\textbf{OpenFOAM-v2412}} \\
			\hline
			& \multicolumn{3}{|c|}{\textbf{\# GPUs}}  & \multicolumn{3}{|c|}{\textbf{\# GPUs}} & \multicolumn{3}{|c|}{\textbf{\# CPUs}}  \\
			\hline
			& \textbf{2} & \textbf{4} & \textbf{8}  & \textbf{2} & \textbf{4} & \textbf{8} & \textbf{200} & \textbf{400} & \textbf{800} \\
			\hline
			\cellcolor[HTML]{DADADA} \textbf{\# cells} & 65.3M & 65.3M & 65.3M & 65.3M & 65.3M & 65.3M & 65.3M & 65.3M & 65.3M \\
			\cellcolor[HTML]{DADADA} \textbf{\# cells/process} & 32.7M & 16.3M & 8.17M & 32.7M & 16.3M & 8.17M & 327k & 163k & 81.7k \\
			\hline \rowcolor[HTML]{DADADA}
			\cellcolor[HTML]{DADADA} \textbf{T110-T10 [s]} & \textbf{302.8} & \textbf{172.41} & \textbf{116.56} & \textbf{341.93}	& \textbf{187.13} & \textbf{108.14} & \textbf{824.07} & \textbf{391.6} & \textbf{227.93} \\
			\hline
			\cellcolor[HTML]{DADADA} \textbf{SIMPLE loop [s]} & 3.068 & 1.748 & 1.195 & 3.458 & 1.894 & 1.093 & 8.347 & 3.994 & 2.326 \\
			\cellcolor[HTML]{DADADA} \textbf{U assembly [s]} & 0.690 & 0.377 & 0.208 & 0.692 & 0.375 & 0.206 & 0.947 & 0.413 & 0.232 \\
			\cellcolor[HTML]{DADADA} \textbf{U predictor [s]} & 0.038 & 0.022 & 0.013 & 0.039 & 0.022 & 0.013 & 0.099 & 0.029 & 0.014 \\
			\cellcolor[HTML]{DADADA} \textbf{U solve [s]} & 1.095 & 0.564 & 0.338 & 1.102 & 0.547 & 0.316 & 4.180 & 1.990 & 1.115 \\
			\cellcolor[HTML]{DADADA} \textbf{P assembly [s]} & 0.065 & 0.040 & 0.024 & 0.064 & 0.039 & 0.023 & 0.149 & 0.057 & 0.043 \\
			\cellcolor[HTML]{DADADA} \textbf{P solve [s]} & 0.165 & 0.124 & 0.251 & 0.553 & 0.304 & 0.181 & 0.738 & 0.527 & 0.329 \\
			\cellcolor[HTML]{DADADA} \textbf{Turbulence [s]} & 0.806 & 0.491 & 0.284 & 0.802 & 0.482 & 0.280 & 1.786 & 0.777 & 0.465 \\
			\hline
		\end{tabular}
	}
	\caption{Detailed section breakdown of strong scaling performances of \myof{}, with AMGX and GAMG solver for pressure, compared with OpenFOAM-v2412, on Leonardo HPC using the occDrivAer test case.}
	\label{tab:leo_strong_scalability}
\end{table}

%% --- LUMI Weak scaling ---%%
\begin{table}[h!]
	\centering
	\resizebox{1.\textwidth}{!}
	{
		\begin{tabular}{|l|c|c|c|c|c|c|}
			\hline
			& \multicolumn{3}{|c|}{\textbf{SPUMA}} & \multicolumn{3}{|c|}{\textbf{OpenFOAM-v2412}} \\
			\hline
			& \multicolumn{3}{|c|}{\textbf{\# GPUs}} & \multicolumn{3}{|c|}{\textbf{\# CPUs}}  \\
			\hline
			& \textbf{4} & \textbf{8} & \textbf{20} & \textbf{400} & \textbf{800} & \textbf{1600} \\
			\hline
			\cellcolor[HTML]{DADADA} \textbf{\# cells} & 65.3M & 121M & 236M & 65.3M & 121M & 236M \\
			\cellcolor[HTML]{DADADA} \textbf{\# cells/process} & 16.3M & 15.1M & 11.8M & 163k & 151k & 147k \\
			\hline
			\rowcolor[HTML]{DADADA}
			\cellcolor[HTML]{DADADA} \textbf{T110-T10 [s]} & \textbf{250.18} & \textbf{244.66} & \textbf{216.5} & \textbf{394.7} & \textbf{374.96} & \textbf{379.57} \\
			\hline
			\cellcolor[HTML]{DADADA} \textbf{SIMPLE Loop [s]} & 2.535 & 2.494 & 2.238 & 3.982 & 3.820 & 3.869 \\
			\cellcolor[HTML]{DADADA} \textbf{U assembly [s]} & 0.531 & 0.476 & 0.367 & 0.745 & 0.680 & 0.710 \\
			\cellcolor[HTML]{DADADA} \textbf{U predictor [s]} & 0.032 & 0.028 & 0.022 & 0.054 & 0.057 & 0.025 \\
			\cellcolor[HTML]{DADADA} \textbf{U solve [s]} & 0.901 & 0.901 & 0.734 & 1.207 & 1.160 & 1.165 \\
			\cellcolor[HTML]{DADADA} \textbf{P assembly [s]} & 0.051 & 0.043 & 0.038 & 0.107 & 0.082 & 0.076 \\
			\cellcolor[HTML]{DADADA} \textbf{P solve [s]} & 0.165 & 0.323 & 0.490 & 0.412 & 0.482 & 0.539 \\
			\cellcolor[HTML]{DADADA} \textbf{Turbulence [s]} & 0.693 & 0.581 & 0.471 & 1.156 & 1.067 & 1.062 \\
			\hline
		\end{tabular}
	}
	\caption{Detailed section breakdown of weak scaling performances of \myof{}, with GAMG solver for pressure, compared with OpenFOAM-v2412, on LUMI HPC system using the occDrivAer test case.}
	\label{tab:lumi_weak_scalability}
\end{table}

%% --- LUMI strong scaling ---%%
\begin{table}[h!]
	\centering
	\resizebox{1.\textwidth}{!}
	{
		\begin{tabular}{|l|c|c|c|c|c|c|}
			\hline
			& \multicolumn{3}{|c|}{\textbf{SPUMA}} & \multicolumn{3}{|c|}{\textbf{OpenFOAM-v2412}} \\
			\hline
			& \multicolumn{3}{|c|}{\textbf{\# GPUs}} & \multicolumn{3}{|c|}{\textbf{\# CPUs}}  \\
			\hline
			& \textbf{2} & \textbf{4} & \textbf{8} & \textbf{200} & \textbf{400} & \textbf{800} \\
			\hline
			\cellcolor[HTML]{DADADA} \textbf{\# cells} & 65.3M & 65.3M & 65.3M & 65.3M & 65.3M & 65.3M \\
			\cellcolor[HTML]{DADADA} \textbf{\# cells/process} & 32.7M & 16.3M & 8.17M & 327k & 163k & 81.7k \\
			\hline \rowcolor[HTML]{DADADA}
			\cellcolor[HTML]{DADADA} \textbf{T110-T10 [s]} & \textbf{432.42} & \textbf{250.18} & \textbf{165.54} & \textbf{852.56} & \textbf{394.7} & \textbf{173.3} \\
			\hline
			\cellcolor[HTML]{DADADA} \textbf{SIMPLE Loop [s]} & 4.376 & 2.535 & 1.691 & 8.605 & 3.982 & 1.758 \\
			\cellcolor[HTML]{DADADA} \textbf{U assembly [s]} & 0.951 & 0.531 & 0.298 & 1.551 & 0.745 & 0.358 \\
			\cellcolor[HTML]{DADADA} \textbf{U predictor [s]} & 0.055 & 0.032 & 0.018 & 0.154 & 0.054 & 0.027 \\
			\cellcolor[HTML]{DADADA} \textbf{U solve [s]} & 1.680 & 0.901 & 0.549 & 2.594 & 1.207 & 0.508 \\
			\cellcolor[HTML]{DADADA} \textbf{P assembly [s]} & 0.080 & 0.051 & 0.030 & 0.210 & 0.107 & 0.046 \\
			\cellcolor[HTML]{DADADA} \textbf{P solve [s]} & 0.228 & 0.165 & 0.303 & 0.937 & 0.412 & 0.185 \\
			\cellcolor[HTML]{DADADA} \textbf{Turbulence [s]} & 1.122 & 0.693 & 0.396 & 2.511 & 1.156 & 0.494 \\
			\hline
		\end{tabular}
	}
	\caption{Detailed section breakdown of strong scaling performances of \myof{}, with GAMG solver for pressure, compared with OpenFOAM-v2412, on LUMI HPC system using the occDrivAer test case.}
	\label{tab:lumi_strong_scalability}
\end{table}

A similar trend can also be observed in the weak scaling plots. 
\myof{} employing AmgX on Leonardo has an efficiency no lower than 90\% in the range of 4-20 GPUs, while the scalability of \myof{} and OpenFOAM-v2412 using GAMG decreases similarly to about 75\% at 20 GPUs. On LUMI, the behavior of these two solvers is better in both cases, with an excellent efficiency of 95\% of OpenFOAM-v2412 at 20 GPUs and about 85\% of \myof{}.

\subsection{Energy consumption}

Most of the results presented in section \ref{sec::performance_evaluation} are focused on performance metrics regarding the time-to-solution of simulations. However, energy consumption and sustainability are becoming increasingly relevant in the HPC sector.
For this reason, we have decided to include a preliminary study that focuses on comparing the energy consumption of OpenFOAM simulations run on CPUs and GPUs.

Three cases have been selected from the ones presented in the previous paragraphs:
\begin{itemize}
	\item \textit{occDrivAer65M, 2 GPUs, GAMG:} considering that GAMG simulations are still far from optimal in terms of strong scalability, this case can be considered as the most efficient GPU case presented in this work.
	\item \textit{occDrivAer65M, 4 GPUs, GAMG:} this case is slightly less efficient than the previous one, when considering strong scalability trends. However, this is characterized by lower time to solution when compared to the CPU reference.
	\item \textit{occDrivAer65M, 800 cores, GAMG:} the fastest case on the CPU side.
\end{itemize}
The selected cases were run while monitoring the energy consumption of the computing nodes. This analysis was conducted exclusively on Leonardo as it provided easier access to power monitoring tools, being the primary HPC system at the authors' institution.
Hardware power usage has been measured employing \textbf{perf}, a Linux performance analyzing tool, for the CPU energy contribution and \textbf{nvidia-smi}, the NVIDIA built-in System Management Interface (SMI), for the GPU energy contribution.

An important remark regards the time-to-solution of the considered cases: the CPU one is approximately 25\% faster than the 2 GPUs case. However, in this comparison, 1 GPU is equivalent to approximately 301 CPU cores. On the other hand, the case running on 4 GPUs is 24\% faster than the CPU reference, and 1 GPU is equivalent to 264 CPU cores. This is translated into the results presented in Table \ref{tab:energy_consumption}.

%%%%%%%%%%%%%%%%%%%%%%%%%%%%%%%%%%%%%%%%%%%%%%%%%%%%%%%%%%
\begin{table}[ht]
	\centering
	\resizebox{1.\textwidth}{!}
	{
		\begin{tabular}{c c c c c c c}
			\hline
			& \# of devices & Energy [Wh] & TDP [W]  & Avg. Power [W] & Load \\
			\hline
			OpenFOAM-v2412 & 16 CPUs (800 cores) & 281.81 & 350 & 346.5 & 99\% \\
			\hline
			\multirow{2}{*}{\myof{}} &   2 GPUs           & 43.21  & 440   & 258   & 59\% \\
			&   1 CPU (32 cores) & 10.67  & 250   & 127.3 & 51\% \\
			\hline
			\multirow{2}{*}{\myof{}} &   4 GPUs           & 44.12  & 440   & 232  & 53\% \\ % ERROR submitted with 42%
			&   1 CPU (32 cores) & 6.54  & 250 & 137.6 & 55\% \\
			\hline
		\end{tabular}
	}
	\caption{CPU and GPU energy consumptions of 100 SIMPLE iterations of the occDrivAer65M test case. TDP refers to Thermal Design Power, the maximum amount of heat a chip can produce under typical heavy use.}
	\label{tab:energy_consumption}
\end{table}
%%%%%%%%%%%%%%%%%%%%%%%%%%%%%%%%%%%%%%%%%%%%%%%%%%%%%%%%%%

Both GPU cases show a combined energy consumption (both the accelerators and the host CPU node are considered) that is significantly lower than the CPU counterpart. Quantitatively, running the case allows users to save more than 80\% energy compared to the original CPU execution. While it is clear that CPUs and GPUs can be characterized by comparable execution times in medium-to-low-sized simulations (e.g., the 65M elements one presented here), switching to GPUs has a huge impact on energy savings. On the other hand, GPU execution should represent a better choice for bigger analyses, considering both time-to-solution and energy consumption. It is interesting to highlight that the 4 GPUs case is not only faster, but also more efficient than the 2 GPUs one when considering the energy consumption. 

\myof{} currently does not include specific optimizations that target GPUs, which could be included soon. In our opinion, this consideration adds even more value to the energy profiling results presented in this section. On the other hand, LUMI's EPYC Milan CPUs are both faster and better in scaling than Leonardo's Sapphire Rapids: a complete evaluation of the combined energy and time-to-solution performances would require a wider set of testing hardware and an extensive study, which is beyond the scope of the current work.
%% ========================================================== %%

%% ========================================================== %%
% CONCLUSIONS
%% ========================================================== %%
\section{Conclusions and future work}
\label{sec::conclusions_and_future_work}

The software presented in this paper represents a crucial innovation in the CFD landscape, enabling the execution of the most adopted open-source simulation framework on modern heterogeneous hardware, characterized by a mix of CPU and GPU accelerators.
The main features introduced in this work can be summarized in three points:

\begin{itemize}
	\item \textit{Transparent memory management:} The introduction of a memory pool simplifies the complex interactions involved in memory management between the CPU and the associated accelerators. The development of an interface composed of a few APIs dedicated to memory-related procedures allows all OpenFOAM developers to tackle the challenge of porting new portions of the code to accelerators, even if they are not experts in GPU programming. This is a significant advantage, as one of OpenFOAM's recognized strengths is the open source community contributing to its development. 
	\item \textit{Compatible programming model:} The implementation of a portable programming abstraction has enabled the isolation of GPU kernels in few dedicated classes. Similar to memory management, this approach simplifies the porting procedure. Additionally, it greatly improves maintainability and potential upgrades of the code, since the introduction or deprecation of backends affects only a few classes.
	\item \textit{Incremental programming model:} If the memory pool is allocated with managed memory technologies, the only effect of an incomplete porting procedure is the presence of slowdowns due to memory page migrations, while the simulation still provides correct results. This allows new features to be introduced incrementally, with profiling tools used to identify slowdowns and replace code sections that cause migrations.
\end{itemize}

These three features allow us to consider \myof{} as a development sandbox, enabling both official maintainers and third-party developers to use it as a starting point for GPU-porting operations.

The findings of this article are not limited to the software features described above. The potential of \myof{} has been demonstrated by simulating a recognized industrial test case, the DrivAer, on two major European supercomputers, Leonardo and LUMI.
Some highlights of the DrivAer tests include:
\begin{itemize}
	\item On Leonardo, one Nvidia A100 is equivalent to between 200 and 300 Intel Sapphire Rapids cores, resulting in excellent energy performance: more than 80\% of energy is saved when running on GPUs. On LUMI, one AMD MI250x is equivalent to 100-150 AMD Epyc Milan cores. We can expect to see better performance in the near future , driven by the rapid evolution of GPU technology stimulated by the  growth of the AI market.
	\item Better scaling behavior for both OpenFOAM-v2412 and \myof{} was observed on LUMI compared to Leonardo. Still, the OpenFOAM GAMG implementation introduces some penalties when executed on GPUs, on both supercomputers. This is confirmed by the higher efficiency obtained with NVIDIA AmgX, which performs considerably better in weak and strong scalability analyses on Leonardo.
\end{itemize}

These results were achieved without adapting any of OpenFOAM's algorithms and data structures to the characteristics of GPU hardware. Future developments could focus on specific improvements to the multigrid solver to reach further performance gains. These could include enhancements to coarsening algorithms, parallel smoothers, and MPI communications. Other important activities could involve loop-fusion techniques, caching of data and matrices, and, in general, favoring operations that increase computing intensity.

For horizontal developments---extending the number of OpenFOAM features that support GPU execution---we plan the inclusion of compressible solvers and domain interfaces, as well as multiphase and heat transfer solvers, which are currently not supported. Assuming similar hardware and a comparable level of GPU memory oversubscription, similar performance gains to those found for the DrivAer test case can  be expected for these solvers as well. In addition, \myof{} can be considered a developer-friendly project, meaning that the wide international OpenFOAM community will play an important role in further extending the amount of GPU-supported functionalities.
%% ========================================================== %%

%% ========================================================== %%
% FUNDING
%% ========================================================== %%
\section{Funding and acknowledgments}

We acknowledge the EuroHPC Joint Undertaking for awarding us access to 
Leonardo at Cineca, Italy, and LUMI at CSC, Finland, under allocation EHPC-DEV-2025D01-014.

This work was co-funded by EuroCC 2. EuroCC 2 has received funding from the European High-Performance Computing Joint Undertaking (JU) under Grant Agreement No. 101101903 and from the Ministry of Enterprises and Made in Italy Progn.P/050001/03/X65 under the National Recovery and Resilience Plan (Mission 4, Component 2 – Investment 2.2 – Partnerships for Research and Innovation – Horizon Europe). The JU is supported by the European Union’s Digital Europe Programme and by Germany, Bulgaria, Austria, Croatia, Cyprus, Czech Republic, Denmark, Estonia, Finland, Greece, Hungary, Ireland, Italy, Lithuania, Latvia, Poland, Portugal, Romania, Slovenia, Spain, Sweden, France, the Netherlands, Belgium, Luxembourg, Slovakia, Norway, Türkiye, the Republic of North Macedonia, Iceland, Montenegro, Serbia, Bosnia and Herzegovina.

This research was conducted as part of the exaFOAM Project, which has received funding from the European High-Performance Computing Joint Undertaking (JU) under grant agreement No 956416.The JU received support from the European Union’s Horizon 2020 research and innovation programme and France, Germany, Spain, Italy, Croatia, Greece, and Portugal.
%% ========================================================== %%

%% References


\bibliographystyle{elsarticle-num}
\bibliography{biblio}

%% ========================================================== %%
% APPENDIX
%% ========================================================== %%
\appendix
\section{simulation setup}
\label{appendix:simulation_setup}
The linear algebra solvers used in the validation runs are described in Table \ref{tab:validation_alg_setup}.
\begin{table}[h!]
	\centering
	\begin{tabular}{c c c c c}
		\hline
		Equation & solver & preconditioner/smoother & tolerance & relative tolerance  \\
		\hline
		p & GAMG & Richardson & 1e-9 & 1e-3 \\
		p & PCG & diagonal & 1e-9 & 1e-3 \\
		p & AmgX & Jacobi & 1e-9 & 1e-3 \\
		U, k, $\omega$ & PBiCG & aDILU & 1e-8 & 1e-3 \\
		\hline
	\end{tabular}
	\caption{Linear solvers setup - validation runs.}
	\label{tab:validation_alg_setup}
\end{table}
Listing \ref{code::val_fvSolution} contains the setup of linear algebra solvers, the SIMPLE algorithm entries, and the relaxation factors (fvSolution dictionary in OpenFOAM). Listing \ref{code::val_fvSchemes} contains the discretization schemes (fvSchemes dictionary in OpenFOAM).

\begin{lstlisting}[language=C++, caption= Validation test case - fvSolution., label=code::val_fvSolution]
	solvers
	{
		
		pcgAmgXJacobi
		{
			solver          AmgX;
			matrixType      CSR;
			dataLocation device;
			mode dDDI;
			AmgXconfig
			{
				config_version 2;
				determinism_flag 1;
				solver
				{
					preconditioner
					{
						print_grid_stats 0;
						print_vis_data  0;
						solver  AMG;
						smoother
						{
							scope  smootherSolver;
							solver BLOCK_JACOBI;
							monitor_residual  0;
							print_solve_stats  0;
							relaxation_factor 0.75;
						}
						algorithm  AGGREGATION;
						selector SIZE_8;
						min_coarse_rows  20;
						print_solve_stats  0;
						aggressive_levels  2;
						presweeps  1;
						interpolator  D2;
						max_iters  2;
						monitor_residual  0;
						store_res_history  0;
						scope  precondSolver;
						max_levels 6;
						cycle  V;
						postsweeps 2;
						coarsest_sweeps 4;
					}
					solver  PCG;
					print_solve_stats  0;
					obtain_timings  0;
					max_iters  500;
					monitor_residual  1;
					store_res_history 1;
					convergence  COMBINED_REL_INI_ABS;
					scope  mainSolver;
					alt_rel_tolerance 0.001;
					tolerance  1e-09;
					norm  L1_SCALED;
				}
			}
		}
		
		pcgDiag
		{
			solver PCG;
			preconditioner   diagonal;
			tolerance        1e-9;
			relTol           0.001;
			maxIter		     3000;
			minIter          1;
		}
		
		pGAMG
		{
			solver GAMG;
			smoother Richardson;
			tolerance        1e-09;
			relTol           0.001;
			maxIter         300;
			minIter          1;
			
		}
		
		p
		{
			$pGAMG;
			//$pcgDiag;
			//$pcgAmgXJacobi;
		}
		
		"(U|k|omega|nuTilda)"
		{
			solver           PBiCG;
			preconditioner   aDILU;
			tolerance        1e-08;
			relTol           0.001;
			minIter          1;
			maxIter          1000;
		}
		
	}
	
	SIMPLE
	{
		residualControl
		{
			p	1e-02;
			U	1e-03;
			"(k|omega|epsilon)" 1e-03;
		}
		nNonOrthogonalCorrectors 0;
		consistent      yes;
		pRefCell        0;
		pRefValue       0;
	}
	
	
	relaxationFactors
	{
		fields
		{
			"(p|pFinal)"               0.3;
		}
		
		equations
		{
			"(U|UFinal)"               0.7;
			"(k|omega|nuTilda)"        0.7;
			"(k|omega|nuTilda)Final"   0.7;
		}
	}
	
\end{lstlisting}

%____________ schemes______________________________%

\begin{lstlisting}[language=C++, caption= Validation test case - fvSchemes, label=code::val_fvSchemes]
	ddtSchemes
	{
		default         steadyState;
	}
	
	gradSchemes
	{
		default         Gauss linear;
		limited         cellLimited Gauss linear 1;
		grad(U)         $limited;
		grad(k)         $limited;
		grad(omega)     $limited;
	}
	
	divSchemes
	{
		default         none;
		div(phi,U)      bounded Gauss linearUpwind grad(U);
		div(U)      Gauss linear;
		turbulence     bounded Gauss upwind;
		div(phi,k)      $turbulence;
		div(phi,omega)  $turbulence;
		
		div((nuEff*dev2(T(grad(U))))) Gauss linear;
	}
	
	laplacianSchemes
	{
		default         Gauss linear corrected;
	}
	
	interpolationSchemes
	{
		default         linear;
	}
	
	snGradSchemes
	{
		default         corrected;
	}
	
	wallDist
	{
		method          meshWave;
	}
	
\end{lstlisting}

The setup used in scalability tests can be retrieved from \cite{occ_drivaer},
except for the linear solvers (Table \ref{tab:scaling_alg_setup}), and the convective discretization scheme (\verb|linearUpwind| in place of \verb|linearUpwindV|).

\begin{table}[ht]
	\centering
	\begin{tabular}{c c c c c}
		\hline
		Equation & solver & preconditioner/smoother & tolerance & relative tolerance  \\
		\hline
		p & GAMG & Richardson & 1e-7 & 0.1 \\
		U, k, $\omega$ & PBiCG & aDILU & 1e-6 & 0.1 \\
		\hline
	\end{tabular}
	\caption{Linear solvers setup - scalability tests.}
	\label{tab:scaling_alg_setup}
\end{table}
%% ========================================================== %%






% \end{thebibliography}
z

\end{document}